\documentclass[aps,
groupedaddress,floatfix]{revtex4}
\usepackage[dvips]{epsfig}

\begin{document}

\title{{\it Ab initio} energy landscape of LiF clusters}

\author{K. Doll$^*$, J. C. Sch\"on, M. Jansen}
\affiliation{Max Planck Institute for Solid State Research, Heisenbergstr. 1, 
D-70569 Stuttgart, Germany}

\date{\today}

\begin{abstract}
A global search for possible LiF cluster structures is performed,
up to (LiF)$_8$. The method is based on
simulated annealing, where all 
the energies are evaluated on the {\em ab initio}
level. In addition, the threshold algorithm is employed to determine the 
energy barriers for the transitions among these structures,
for the cluster (LiF)$_4$, again
on the {\em ab initio} level; and the corresponding tree graph is
obtained.
\end{abstract}

\email{k.doll@fkf.mpg.de}

\maketitle

\section{Introduction}

The question of the possible
structures of a chemical system, both in the case of molecules and of solids,
is of huge importance
\cite{Cohen89,Catlow90,Schoen96b,Jansen02b,Woodley2008,Ferrando2008,Schoen2010pss,WalesBuch}.
Identifying these structures requires
the determination of the local minima on the energy landscape of the system, 
or more generally the locally
ergodic regions (basins). At low temperatures, individual local minima,
that are surrounded by sufficiently high energy barriers, are kinetically
stable. There are numerous studies of clusters, where global 
searches for their minima have been performed.
This is a non-trivial task, as the number of minima is expected to 
increase exponentially with system size, see e.g. 
\cite{Berry1993,Stillinger1999}. 

However, finding the local minima is only one part of identifying promising
structures. The second task is the analysis of the barrier structure of the
landscape, in particular the determination of the energetic barriers
separating the local minima, which control the stability of the structure.
To describe the energy landscape and barriers of a system,
tree graphs \cite{Hoffmann1988,Schoen1996,Becker1997,WalesBuch} 
can be employed. Examples for such tree graphs for solids
can be found in 
\cite{Schoen1996,Wevers1999}, and for clusters in \cite{WalesBuch},
respectively.

In the present work, we have chosen LiF clusters as the system to be
studied, since alkali halide clusters have served as important model systems
for complex energy landscapes. Such clusters
have previously been investigated with various methods. In early calculations,
atoms were placed in random positions and then moved (to be more
precise, a quench was performed, i.e. a move is suggested and accepted, 
if it lowers the energy) \cite{Martin1983}. A different
scheme applied for NaCl clusters \cite{Phillips1990}
was to build up clusters by adding one atom at a time:
an atom is added in a random location, 
then the energy of the new structure is computed,
and according to an acceptance probability, this site is accepted or not.
When the requested number of atoms has been added and the intended cluster
size is obtained, then subsequently a gradient minimization of the cluster is
performed. With a sufficiently high number of trials, this is supposed
to give a good overview of the energy landscape. A further technique
used was simulated annealing with molecular dynamics or simulated
annealing where the system is represented by a
Gaussian density distribution for each particle \cite{Amara1996}.
Basin hopping was applied \cite{Doye1999} to charged clusters (NaCl)$_n$Cl$^-$,
and to the related system Al$_n$N$_n$ \cite{Costales2005}.
In addition, a genetic algorithm has been applied to NaCl clusters
\cite{Michaelian1998,Kabrede2002}.

The aforementioned studies employed model potentials. A
comprehensive
{\it ab initio} study on the level of Hartree-Fock, second order perturbation
theory and the coupled pair functional has been performed for selected
clusters \cite{Ochsenfeld1992}. Moreover, {\em ab initio}
molecular dynamics simulation
was applied to small NaCl clusters (stoichiometric and non-stoichiometric)
\cite{Barnett1995}. {\it Ab initio}
calculations on known structures have nowadays
become routine work, see e.g. \cite{Lintuluoto2001,Lai2002}.
Among the studies mentioned so far, most apply to alkali halides in
general, or to NaCl clusters as a special case. LiF clusters are addressed
in references \cite{Aguado1997,Lintuluoto2001,Fernandez2009}. 
For a review on the molecular
structure of metal halides, see \cite{Whetten1993,Hargittai2000}, and on the
modeling of clusters, see \cite{Catlow2010}.

The search for local minima
is commonly split in two parts \cite{Schoen1996}: a global search for 
structure candidates, and a subsequent local
optimization with accurate {\em ab initio} energy calculations.
Usually, the global search is 
performed with empirical potentials, because the
global search is the time-consuming part, and empirical potentials
require little CPU time. In contrast, in
the present work, both steps, i.e. global search and local optimization
are performed on the {\em ab initio} level. The procedure is still split in
two steps: first, the global optimization is performed with
{\em ab initio} calculations on a lower level of accuracy which makes
them much faster, however. Nevertheless, this is a non-trivial task as 
the CPU times are still much higher than with empirical potentials.
The local optimization is then performed
with the usual high accuracy, as this step requires less CPU time than
the global optimization.
Employing {\em ab initio} methods in both steps of the
search has several advantages: the most important is
that {\em ab initio} methods work for general systems, whereas 
empirical potentials work reasonably well for ionic systems, but less so
for covalent systems; and some knowledge of the expected bond type is 
required in advance to choose the potential. 

Recently, 
we demonstrated, in the context of the structure prediction of solids, that
a full {\it ab initio} treatment is feasible in both stages, i.e. the global
search and the subsequent local optimization can both be performed on
the {\it ab initio} level. LiF was chosen as an example for an ionic solid
\cite{DollPCCP}, and BN as an example for a covalent solid \cite{DollBN}.
In the case of lithium fluoride, 
it was found that the relevant minima
were the same when full {\it ab initio} structure prediction was performed,
compared to the earlier studies with model potentials
\cite{Schoen95} 
(for a brief summary see also \cite{Dollfuerpisani}).
Concerning clusters, the earliest work which combined simulated annealing
with {\it ab initio} energies employed the Car-Parrinello method 
\cite{CP1985}, and was performed for e.g. silicon \cite{Andreoni} and
selenium \cite{Hohl1987} clusters. 
In order to include the nuclear motion explicitly,
{\it ab initio} path integral molecular
dynamics \cite{Marx1996,Cheng1995,Tuckerman1997} was suggested as an
extension. In these methods, the atom moves are determined by
molecular dynamics.
Simulated annealing in combination with
stochastic moves and {\it ab initio} energies had been applied earlier
to various clusters, e.g. the clusters Li$_5$H \cite{KeshariCPL1994},
Mg$^+$(H$_2$O)$_n$ \cite{Asada1996} or lithium clusters \cite{Srinivas2000}.

The goal 
of the present article is the study of the energy landscape of LiF
clusters, completely on the {\em ab initio} level. 
The first step is to study the feasibility of 
a global optimization with
simulated annealing and employing {\em ab initio} energies
for these systems, and to
compare with earlier studies as far as available. 
The second and more challenging step is to determine
the energy barriers between the structures, again
on the {\em ab initio} level. The latter task is fairly time-consuming, and had
not been attempted for the solids which had been studied previously with
{\it ab initio} simulated annealing. The intention of the present article
is to test and demonstrate the feasibility of the latter step. LiF
has been chosen, since the binding situation
is fairly ionic, and thus convergence for unknown structures
is relatively easy to achieve. Thus, 
comparatively simple {\it ab initio} methods
such as Hartree-Fock or density functional theory are expected to work
reasonably well. The outcome of the present study, and also
the information about the demand on computational resources, is important 
in order to
estimate the effort for more difficult systems with a partially
covalent character, which will be future targets. 

In the following paragraph, the
technical details will be described. Then, the results are presented
and discussed; and finally the article will be summarized.

\section{Method}
\label{methodsection}

As mentioned above, the general optimization procedure
consists of several steps:
first, a simulated annealing run with a subsequent stochastic quench
is performed, to identify possible
candidate structures. This is followed by a local optimization based
on analytical gradients. 
This is repeated many times, in order to identify as large as possible
a set of
structure candidates, and to obtain some statistics about the structures
found.

The details for the LiF calculations are described in the following
paragraphs.
Between 1 and 8 formula units of LiF,
i.e. from 1 lithium and 1 fluorine atom
up to 8 lithium and 8 fluorine atoms were placed at random positions
in a large box. The box
was initially cubic with a size corresponding to three times the total volume
of the atoms/ions
as estimated from the atomic/ionic radii. Note that no periodicity is
used, i.e. the box is not repeated in space. The box is mainly
used to place the atoms initially, and, in general,
moves are defined which allow
to change the box size. In the case of molecules, one might consider
not allowing such moves and keep the size of the box fixed, 
which might, however, lead to the search  requiring 
more steps to reach a local minimum.

The length of the simulated annealing run was in the range between
5000, for the smaller clusters, and 75000 steps, for the largest clusters,
respectively.
The initial temperature was chosen in the range of 1-10 eV, (1 eV
corresponds to 11604 Kelvin),
and was reduced by up to $\sim 50 \%$ during the simulation, 
in the longer runs. 
This temperature is used
when performing the Metropolis
Monte-Carlo simulation: the energy difference per atom $\Delta E$
is divided
by this temperature when evaluating the expression $exp(-\Delta E/k_BT)$.

The simulated annealing was
followed by a quench with 10000 steps, i.e.\ a simulated annealing run
with a temperature of 0 eV, which means that only downhill moves are allowed
during the quench.

The moves were chosen as: moving individual atoms (70\%), exchanging
atoms (10\%), changing the size of the simulation box (20\%). Note that
exchanging atoms is a useful move especially if the initial random
structure has neighboring pairs of the type Li-Li or F-F. 
No symmetry was prescribed during the simulated annealing and quench runs,
i.e. the point group was always $C_1$.

A minimum distance between two atoms (given by
the sum of the radii of the atoms, multiplied
by 0.7) was prescribed in order
to avoid unrealistic geometries which may lead to
numerical
instabilities. The radii used were based on tabulated values
for atomic and ionic radii, as a function of charge, and the 
Mulliken charge computed for the previous configuration.
In those moves which change the size of the simulation box,
the probability of reducing the lattice constant was enlarged to 70\%,
to speed up the reduction of the cell size.

With this choice of the parameters, one simulated annealing and
quench run lasted from minutes for the smaller clusters
up to 1 day for the largest cluster
considered, i.e. (LiF)$_8$.

The {\it ab initio} 
calculations were performed with the CRYSTAL06 code \cite{Manual06},
which is based on local Gaussian type orbitals. 
The basis sets employed are displayed in table \ref{basissettable}.
Two slightly different basis sets are used during the global search
and the local optimization.
During the global search, slightly tighter $sp$
functions were chosen for the two outermost fluorine exponents, in order 
to enhance the numerical
stability and the speed of the calculations.
In order to significantly reduce the CPU time, the following simplifications
were used during the global search: the thresholds for the integral selection
\cite{Manual06} were reduced from
the default values ($10^{-6}, 10^{-6}, 10^{-6}, 10^{-6}, 10^{-12}$) to 
$10^{-4}, 10^{-4}, 10^{-4}, 10^{-4}, 10^{-8}$, and the convergence threshold
of the self-consistent field cycles was reduced from $10^{-7}$ (default)
to $10^{-3}$.

The global search was performed on the level of Hartree-Fock theory,
i.e.\ all the energies of all the structures appearing during the
simulated annealing and quench were computed from first principles.
As has been demonstrated earlier \cite{DollBN}, Hartree-Fock theory
has the advantage that convergence for random structures is
facilitated due to the large HOMO-LUMO gaps (gaps between highest occupied
and lowest unoccupied molecular orbital) on this level of theory, compared
to the much smaller LDA (local density approximation) gaps.

The local optimization is essentially a routine task and
employed analytical gradients
as implemented in the CRYSTAL06 release 
\cite{IJQC,CPC,KlausDovesiRO,KlausDovesiRO1d2d,Mimmo2001}. 
This local optimization
was performed on the level of Hartree-Fock, and subsequently, the
optimal Hartree-Fock structure was again optimized on the LDA level. 
In very few
cases, a structure found on the Hartree-Fock level transformed
to a different structure on the LDA level. These structures were
dropped, and only those structures were included which remained
the same after the LDA optimization.
This way, a consistent set of local minima was obtained for clusters of the
size (LiF)$_1$ to (LiF)$_8$. The symmetry of these clusters was analyzed
with the program SYMMOL \cite{SYMMOL}.

Going beyond the search for local minima, an exploration of the barrier
structure using threshold runs \cite{Schoen1996} was performed for
the system (LiF)$_4$. A threshold run starts
from a local minimum and explores the part of the energy landscape that
can be reached from this configuration without crossing a given energy lid. 
A new configuration is generated 
(according to a certain move class), and this configuration
is accepted if its energy is below a certain threshold. Then, one
or several quenches are 
performed starting from some of the high energy structures
encountered during the threshold run. When low thresholds are applied,
these quenches tend to return to the starting minimum.
However, they can reach a new structure, if the threshold energy is high
enough to overcome the barrier between the initial structure and the
new one. This way, an upper bound of the energy barrier between
two structures can be determined. 

Here, five threshold runs with a length of 100000 steps each were performed
for each energy threshold and each starting minimum. 
The runs were interrupted after 20000, 40000,
60000, 80000 steps, in order to perform three quenches with 20000 steps each.
This results in a total of 400000 steps per threshold run (100000+5*3*20000).
Note that a quench
is again a stochastic procedure, and thus three quenches can, in principle, end
up in three different local minima. This, however, turned out to be only rarely
the case (in about 1\% of the cases; in all the other cases, the same
minimum structure was obtained from the three quenches).
The thresholds were chosen in an iterative way so that
the energy barriers could be determined to $\sim 0.01$ eV/atom. A more
systematic approach employing a set of threshold values as in 
\cite{Wevers1999}
would also provide more statistical information such as the probability
to find a certain structure as a function of the threshold energy, but
would require significantly more CPU time. 

As the threshold run intends to give a realistic
simulation of the transition from one cluster
structure to another one, the exchange of atoms was not permitted, and
instead, only atom moves (80\%) and changes of the box size (20\%) were
allowed. 

With the present choice of
parameters (employing the set of less accurate parameters, as used in
the global search), 
the CPU time for one energy calculation of a (LiF)$_4$ cluster
is about 1 second on a single CPU (Intel Xeon 5150, 2.66 GHz). More precisely:
in the simulated annealing run,
the initial geometry is like a gas, the atoms are far apart, and
many integrals are discarded (due to the selection criteria of the
integrals). 
Thus, one energy 
calculation at this geometry takes
only a fraction of a second (less than 0.2 seconds). In the final
geometry at the end of the simulated annealing run, 
the atoms are close, and many more integrals have to be
evaluated: as the atoms are closer, the overlap integrals are larger, and
fewer integrals can be neglected.
One energy calculation takes now up to 2 seconds. During the
threshold run, the CPU time for one energy calculation 
is of the order of 1-2 seconds.

One threshold run with 400000 energy calculations
thus lasts of the order of 400000 seconds
(5 days), and thus, with 5 threshold runs 
per lid, about 4 weeks of CPU time are required. This
can obviously be trivially parallalized by employing 5 CPUs, one CPU for
each threshold run. An 
initial threshold may be estimated as e.g. 0.5 eV/atom, and subsequently
this value can be bisected, in order to determine the energy at which
a transition becomes feasible. The bisection is repeated, until
the desired accuracy of the barrier is achieved. 
The information is insofar redundant, as 
a transition between two cluster types can be observed in both directions:
e.g. when the barrier from the cube-shaped cluster to the ring-like cluster
is known, then this can be used to have an initial guess for the
threshold energy which is to be used for the ring-like cluster.

Similarly, the guess of the initial threshold can be better adjusted when
some results from other clusters become available (i.e. when the order
of magnitude of the barrier is known). In total, for
each of the 5 different (LiF)$_4$ clusters,
about 5 threshold values had to be selected, in order to determine
the tree graph. This results in a total of 25*4 weeks of CPU time on
a single CPU for the whole tree graph.

\section{Results and discussion}

\subsection{Structures found}

The most relevant structures found for the (LiF)$_n$ ($n$ = 1,...,8) clusters
are displayed in figures
1-6, 
visualized with XCrysDen \cite{XCrysDen}. 
The statistics are compiled in table \ref{Structuresfound}.
Concerning comparisons with other results for alkali halide clusters,
it should be noted that most of the available literature 
data refers to NaCl clusters, and there is only few
data available explicitly for LiF clusters 
\cite{Aguado1997,Lintuluoto2001,Fernandez2009}. 
However, due to homology, one would expect similar minimum structures
to be present in LiF and e.g. in NaCl clusters.

For $n$ = 1, 2 and 3 formula units, only one structure was found for each
system (figure \ref{1-2-3LiF}) that constituted a stable minimum
for both HF (Hartree-Fock) and LDA calculations. Additional modifications that
were stable only on the HF level appeared one time each for $n$ = 2 and 3.
These structures were thus dismissed.
A double-chain like structure was never observed. This confirms
the statement in \cite{Lintuluoto2001} 
that such a double chain is stable for Na$_3$Cl$_3$ \cite{Ochsenfeld1992},
but not for Li$_3$F$_3$.

For four formula units, four structures were found in 75 simulated
annealing and quench runs, and a fifth during the threshold runs.
These structures agree with those given for (NaCl)$_4$ clusters in
\cite{Ochsenfeld1992}
(apart from the structure in figure 1e of \cite{Ochsenfeld1992} which
was not observed).

For $n$ = 5, seven energetically low-lying structures were found (figure 
\ref{5LiF}). Three of them correspond to the set of minima of (NaCl)$_5$
as published in references \cite{Phillips1990,Ochsenfeld1992}; 
the lowest structure in energy
agrees with the global minimum of (NaCl)$_5$: 
a cube with one additional LiF attached to one edge.
We note that one particularly low-lying structure (5c) 
was found which had not been observed in the earlier work on NaCl clusters
\cite{Phillips1990}.

Figure \ref{6LiF} presents the eight structures with lowest
energy obtained for (LiF)$_6$. 
Structure 6a with two six-membered rings atop of each other
is the energetically most favorable one. 
This structure is a fragment from the 5-5 \cite{Schoen95}
(or 'hexagonal MgO') structure which is a hypothetical low-lying modification
of LiF \cite{DollPCCP}. This result is very interesting for experimental
chemistry, since it suggests that the 5-5-type modification of LiF might be
accessible via the deposition and growth of selected (LiF)$_n$-clusters
on an appropriate substrate (see also \cite{Schoen06ZNF}).
The structure with the second lowest energy is a cuboid cut from
the rock salt structure which is the modification observed 
for bulk LiF at standard conditions. Structure 6c 
consisting of a cube, with a square attached, is ranked number three in energy.
Note that in \cite{Phillips1990,Ochsenfeld1992}, a similar structure is
found where the 3 $\times$ 2 rectangle in 6c is opened up. 
This structure was also found in the present work and is a
candidate structure on the HF level. When optimizing it
on the LDA level, it became the structure 6c, however. Thus,
this additional structure was not included
in the statistics. 
Besides those structures in table \ref{BestStructureEnergies}, additional
minima with higher energies were observed. Furthermore, one
would expect more structures resembling 6d to exist, i.e. consisting of
a cube with two LiF units attached to different edges of the cube.
Finally, a twelve-membered ring structure is now energetically less favorable
than the more compact structures, which is to be expected.

For $n$ = 7, the lowest energy structure found agrees with the
one from the 
literature for (NaCl)$_7$ \cite{Phillips1990,Lintuluoto2001} and (LiF)$_7$ 
\cite{Aguado1997,Lintuluoto2001}. In figure \ref{7LiF}, 
the eight structures found with the lowest energy are shown.
Note that some of the structures
exhibit chirality. In some cases (7c and 7d, 7f and 7g), 
both enantiomers were found in the
simulations, though not in the case of the structure 7b. 
The latter minimum was found only once, and we expect
that increasing the number of runs should also produce the
second enantiomer.
Chirality can thus be used as a test of how thorough the energy landscape
has been searched, since one would expect to find both
enantiomers with the same probability.

The largest clusters investigated contained eight LiF formula units. The lowest
in energy are displayed in figure \ref{8LiF}. The energetically
most favorable structure (8a)
has $S_4$ symmetry, and corresponds to the third lowest energy
isomer of (NaCl)$_8$ \cite{Ochsenfeld1992}. The cuboid
is next in energy and was lowest in energy for (NaCl)$_8$ 
\cite{Ochsenfeld1992}. However, two energetically low-lying minima 
listed in earlier studies
were not observed in the present 
work: an eight-membered double ring with $D_{4d}$
symmetry (for a corresponding NaCl cluster, 
see \cite{Ochsenfeld1992}, and for LiF, see 
\cite{Aguado1997,Lintuluoto2001,Fernandez2009}), 
and a structure made of three cubes, with $C_s$ symmetry 
(for a corresponding NaCl cluster, see \cite{Ochsenfeld1992}, and for
LiF, see \cite{Aguado1997}). 
To check their importance, the energies of these minima were computed,
yielding
 -856.2474 $E_h$ (HF) and -853.8478 $E_h$ (LDA) 
for the structure with $D_{4d}$ symmetry and
 -856.2275 $E_h$ (HF) and -853.8433 $E_h$ (LDA) 
for the structure with $C_s$ symmetry, respectively.
This would put the latter two structures
in the second and fourth place (HF), or third
and fourth place (LDA), respectively. It is expected that these minima
could be found if more simulated annealing runs were performed.

In figure \ref{vergleichenergien} and table \ref{BestStructureEnergies}, 
the energies (per formula unit) of the most favorable
structures are compared.
They decrease monotonously with cluster size
and thus it is always energetically more favorable
to form a bigger cluster from two smaller ones. However, if
one considers the difference in chemical potentials (approximated by
energy differences $\mu_n=E_{n}-E_{n-1}$, 
$\Delta \mu_n=\mu_{n+1}-\mu_n=
(E_{n+1}-E_{n})-(E_{n}-E_{n-1})=E_{n+1}-2E_{n}+E_{n-1}$), one
notes that for certain cluster sizes it is slightly favorable to combine a
(LiF)$_{n+1}$ and a (LiF)$_{n-1}$ cluster to form two (LiF)$_n$ clusters.
This is also displayed in figure \ref{vergleichenergien}: 
$\Delta\mu_n$
is positive for $n=4,6$ and negative for $n=5,7$ which indicates
that the clusters with $n=4,6$ have a higher stability.
The reason appears
to be that structures which resemble 
a fragment from the bulk are more favorable, 
i.e.\ cuboids or distorted cuboids, which cannot be constructed
with 5 or 7 formula units (for a discussion of this issue, see also
\cite{Whetten1993,Woodley2008}). This
is reminiscent of magic cluster numbers in metallic and intermetallic
clusters, but due to the ionic nature of the Li-F bonds, the effect is 
less pronounced.

We note that with increasing size,
rings become less favorable compared
to the other configurations which are more bulk-like. This is to
be expected, as their energy will approximate 
the energy of a one-dimensional chain.

Concerning the statistics, for the larger clusters (i.e. $n$ $\ge$ 5), 
the percentage
of successful runs becomes smaller; i.e. those runs which end up in one
of the lowest energy minima in table 
\ref{Structuresfound}. Quite generally, the structures of the
global minima found usually agree with the ones
known from the literature for LiF, or for NaCl. The only exception is 
(LiF)$_8$, where the minimum found in the present work
is more favorable than the one previously suggested
for (LiF)$_8$ \cite{Aguado1997,Lintuluoto2001,Fernandez2009}.

\subsection{Energy barriers}

The barrier landscape
of the system
(LiF)$_4$ was investigated
with the threshold algorithm \cite{Schoen1996}, on the Hartree-Fock level. 
Note that all these calculations were performed with the parameters
used during the global search, as these calculations are very time-consuming,
and the weaker parameters used during the global search lead to a
significant speed-up. No local optimizations were performed, for reasons
of consistency: during the threshold run, the energies
were computed with less accurate {\it ab initio} parameters (as explained
in \cite{DollPCCP}), and thus
also the energies of the local minima were computed on this level.
The energies displayed in figure \ref{treegraph4LiF} are
thus slightly different from the ones
in table \ref{Structuresfound}.
From the results of the threshold runs, the tree graph for the 
{\em ab initio} energy landscape of (LiF)$_4$ was constructed
(see figure \ref{treegraph4LiF}). 

We note that the cube (4a) and the ring-like structure (4b) have
a relatively high stability and the barriers are larger than
0.1 eV/atom (i.e. 0.8 eV in total). 
A crude estimate of the lifetime of such a cluster can
be performed by applying Arrhenius law $k=Aexp({-E_B/k_BT})$: 
if we assume a prefactor of $A=10^{13}\frac{1}{s}$ (in the range of the
vibrational frequencies, see \cite{Redington1995}), then at a temperature
of 300 K, a barrier $E_B$ of 0.8 eV would result in a rate constant 
$k$ of the order of 0.4 $\frac{1}{s}$. Thus, one would estimate that
both structures might be
observable in the experiment. There is indeed some experimental evidence for
a ring-like (LiF)$_4$ cluster \cite{Redington1995}.
The other three structures found have virtually no energy barrier,
and easily transform to the ring or cube structure. This is to be
expected, as one can see that the rectangular structure (4c) can easily be
deformed to become ring-like (4b). In addition, the two
very high lying minima (4d and 4e)
were so rarely observed during the global search and the threshold
runs, that the basins corresponding to these structures have either
a low energy barrier or a very small volume in configuration space.
This is consistent with the following fact: when threshold
runs were started in structure 4d or 4e, with an energy lid
only slightly above the energy of these minima, then the walkers
only rarely returned to the starting minimum, but instead usually
changed to the ring, cube or rectangular structure.

\section{Conclusion}
Employing simulated annealing as a global optimization method, 
and using {\em ab-initio} energies during both the global and the local
optimization, the structures of lithium fluoride clusters (LiF)$_n$ ($n$ =
1,...,8)
have been predicted. In particular,
the lowest energy structure agreed for all
cluster sizes with, or was more favorable than, 
the known (LiF)$_n$ minimum structures. 
The majority
of the low-lying minima previously known from calculations on
various alkali halide clusters such as (NaCl)$_n$ was found.

As a new methodological development, the threshold algorithm was 
applied to a molecular system, (LiF)$_4$, where all the energy calculations
were performed on the {\it ab initio} level. This way, the
energy barriers separating 
the minima of the energy landscape of this cluster were computed,
and a tree graph representation of the energy landscape of LiF$_4$ was
constructed. Based on this tree graph,
two of the cluster modifications may be accessible in the experiment.

Estimating the stability of the various clusters raises the general 
issue of the accuracy of the calculated barriers separating 
the cluster modifications. Clearly, in the ideal case, 
one would employ highly accurate quantum chemical methods, 
such as coupled-cluster calculations and possibly multi-reference methods. 
However, the determination of the barriers requires
very large computational resources, and is thus not feasible on
this higher level of theory.
Thus, some compromise between accuracy and computational expense must be
established. In the present study, the barriers have been computed on the
Hartree-Fock 
level, where some additional calibration work was necessary, because 
standard Hartree-Fock calculations would be too slow. This type of calculation 
is orders of magnitude more expensive than employing an empirical potential, 
yet still not as accurate as coupled-cluster calculations (see,
e.g. \cite{Lynch2003}). 
Nevertheless, the present work also demonstrates that the global exploration 
of the barrier structure of the energy landscape of small systems is feasible
on the {\em ab initio}
level, though not with the desirable highest accuracy. In particular for 
systems where no trustworthy empirical potentials are available 
that can well describe the barrier structure, such as e.g. compounds 
with a more covalent character, the present approach may well prove 
to be fruitful and significantly improve the quality of the energy 
landscape explorations, in spite of its current limitations.

\clearpage

\begin{table}
\begin{center}
\caption{\label{basissettable}
Basis sets used for the global search (I) and the local optimization
(II).}
\vspace{5mm}
\begin{tabular}{cccccc}
\hline\hline
\multicolumn{2}{c}{basis set I} & \multicolumn{2}{c}{basis set II} \\
exponent & contraction &  exponent & contraction \\
\hline \multicolumn{4}{c}{Li} \\
\multicolumn{4}{c}{$s$}  \\
840.0 & 0.00264 & 840.0 & 0.00264 \\
217.5 & 0.00850 & 217.5 & 0.00850 \\
72.3  & 0.0335  & 72.3  & 0.0335  \\
19.66 & 0.1824  & 19.66 & 0.1824  \\
5.044 & 0.6379  & 5.044 & 0.6379  \\
1.5   & 1.0     & 1.5   & 1.0     \\    
\multicolumn{4}{c}{$sp$}  \\
0.525 & 1.0 1.0 & 0.525 & 1.0 1.0 \\
\hline \multicolumn{4}{c}{F} \\
\multicolumn{4}{c}{$s$}  \\
13770.0 & 0.000877 & 13770.0 & 0.000877 \\
1590.0 &  0.00915  & 1590.0 &  0.00915 \\
326.5 & 0.0486     & 326.5 & 0.0486    \\
91.66 & 0.1691     & 91.66 & 0.1691    \\
30.46 & 0.3708     & 30.46 & 0.3708    \\
11.50 & 0.4165     & 11.50 & 0.4165    \\
4.76  & 0.1306     & 4.76  & 0.1306    \\
\multicolumn{4}{c}{$sp$}  \\
19.0 & -0.1094  0.1244  & 19.0 & -0.1094  0.1244    \\
4.53 & -0.1289  0.5323  & 4.53 & -0.1289  0.5323    \\
1.37 &  1.0     1.0     & 1.37 &  1.0     1.0       \\
\multicolumn{4}{c}{$sp$}  \\
0.45 & 1.0  1.0 & 0.437 & 1.0  1.0\\
\multicolumn{4}{c}{$sp$}  \\
0.20 & 1.0  1.0 & 0.147 & 1.0  1.0 \\
\hline\hline
\end{tabular}
\end{center}
\end{table}
\clearpage

\begingroup
\squeezetable
\begin{table}
\begin{center}
\caption{\label{Structuresfound}The total energies of the various clusters,
in hartree units, and statistics of the outcome of the simulated annealing
runs. A run is counted as successful, if
one of the low lying minima was found.}
\begin{tabular}{cccccc}
\hline\hline
number of &  configu- & HF & LDA & number of \\
formula & ration & & & times found\\
units & & & &  \\ \hline 
1 & a & -106.9444 & -106.6406 & 1/1  \\  \hline 
2 & a & -213.9925 & -213.3883 & 30/30\\  \hline 
3 & a & -321.0393 & -320.1310 & 60/60\\  \hline 
4 & a & -428.0742 & -426.8772 & 41 \\
  & b & -428.0748 & -426.8608 & 23 \\
  & c & -428.0616 & -426.8554 & 10 \\
  & d & -428.0396 & -426.8384 & 1 \\
  & e & -428.0336 & -426.8308 & 0 $^a$ \\
  & \multicolumn {3}{c}{successful runs / total runs:} & 75 / 75 \\ \hline
5 & a & -535.1039 & -533.6054 & 15 \\ 
  & b & -535.0961 & -533.5891 & 1 \\ 
  & c & -535.0979 & -533.5888 & 15 \\ 
  & d & -535.1058 & -533.5862 & 7 \\ 
  & e & -535.0815 & -533.5766 & 2\\ 
  & f & -535.0859 & -533.5765 & 1\\
  & g & -535.0725 & -533.5685 & 1 \\ 
  & \multicolumn {3}{c}{successful runs / total runs:} & 42 / 50 \\ \hline
6 & a & -642.1672 & -640.3706 & 46 \\ 
  & b & -642.1567 & -640.3670 & 2 \\
  & c & -642.1353 & -640.3364 & 6 \\
  & d & -642.1317 & -640.3316 & 4\\
  & e & -642.1360 & -640.3238 & 18 \\ 
  & f & -642.1294 & -640.3221 & 10 \\
  & g & -642.1250 & -640.3172 & 4 \\
  & h & -642.1347 & -640.3101 & 3\\
  & \multicolumn {3}{c}{successful runs / total runs:} & 93/120 \\ \hline
7 & a & -749.1967 & -747.1024 & 6 \\
  & b$^b$ & -749.1856 & -747.0938 & 1\\
  & c & -749.1929 & -747.0932 & 12\\
  & d & -749.1929 & -747.0932 & 11 \\
  & e & -749.1888 & -747.0920 & 4 \\
  & f & -749.1755 & -747.0753 & 2\\
  & g & -749.1755 & -747.0753 & 1\\
  & h & -749.1714 & -747.0700 & 4\\
  & \multicolumn {3}{c}{successful runs / total runs:} & 41 / 80 \\ \hline
8 & a & -856.2524 & -853.8572 & 16 \\
  & b & -856.2383 & -853.8552 & 3 \\
  & c & -856.2216 & -853.8226 & 3\\
  & d & -856.2216 & -853.8226 & 1 \\
  & e & -856.2199 & -853.8181 & 1 \\
  & f & -856.2182 & -853.8159 & 2 \\
  & g & -856.2142 & -853.8152 & 1 \\
  & \multicolumn {3}{c}{successful runs / total runs:} & 27 / 60 \\
\hline\hline
\multicolumn {5}{c}{$^a$ This structure was found only in the threshold run.}\\
\multicolumn {5}{c}{$^b$ The second enantiomer of this chiral structure was not found.} \\
\end{tabular}           
\end{center}
\end{table}
\endgroup

\begin{table}
\begin{center}
\caption{\label{BestStructureEnergies}
The total energies  of the most favorable clusters, per formula unit,
in hartree units, at the Hartree-Fock and LDA level; and
the same quantity for the ring structures.}
\vspace{5mm}
\begin{tabular}{cccc}
\hline\hline
number of &  configu- & HF & LDA  \\
formula & ration & & \\
units & & &   \\ \hline 
1 & a & -106.9444 & -106.6406 \\  
2 & a & -106.9962  & -106.6941\\ 
3 & a & -107.0131 & -106.7103 \\ 
4 & a &  & -106.7193   \\
  & b & -107.0187   & \\
5 & a &  & -106.7211   \\ 
  & d & -107.0212   &  \\ 
6 & a & -107.0279   & -106.7284   \\ 
7 & a & -107.0281   & -106.7289   \\
8 & a & -107.0315   & -106.7322   \\
\hline
Rings \\
3 & a & -107.0131 & -106.7103 \\ 
4 & b & -107.0187   & -106.7152   \\ 
5 & d & -107.0212   & -106.7172   \\
6 & h & -107.0224   & -106.7184 \\
\hline\hline
\end{tabular}           
\end{center}
\end{table}

\clearpage

\begin{widetext}

\begin{figure}
\caption{ (Color online) 
Structures of (LiF)$_n$ clusters with $n$=1,2,3 formula units. 
Lithium is displayed as a small red circle, fluorine
as a large turquoise circle.}
\includegraphics[width=3.cm]{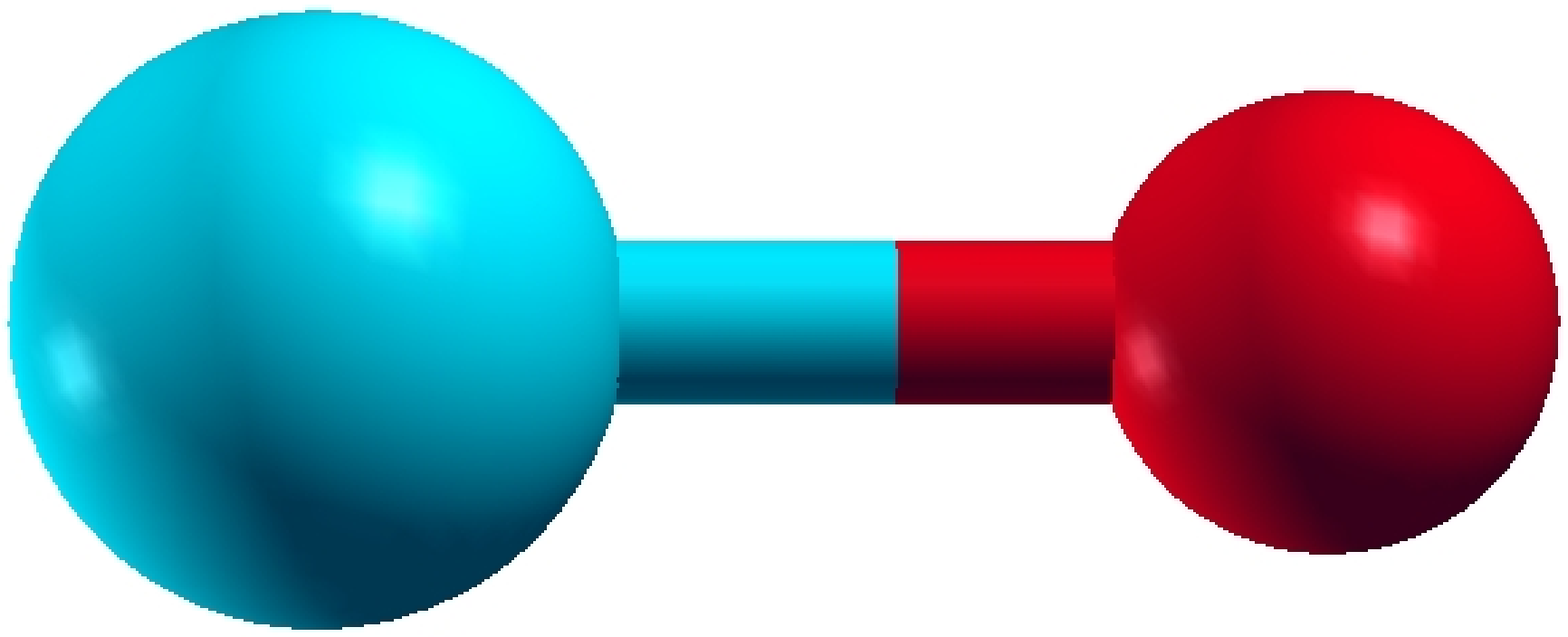} 
\includegraphics[width=3.5cm]{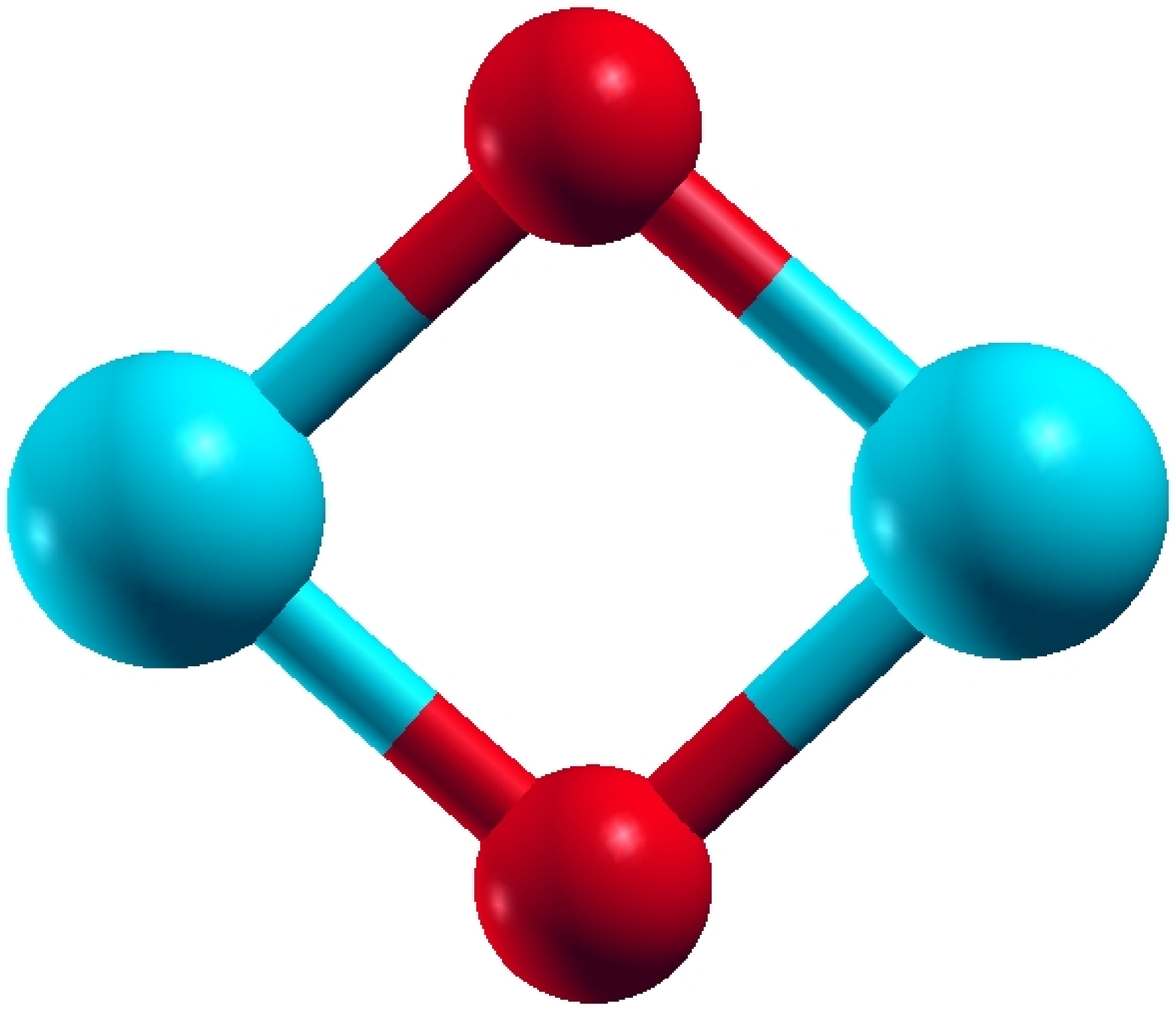} 
\includegraphics[width=4.5cm]{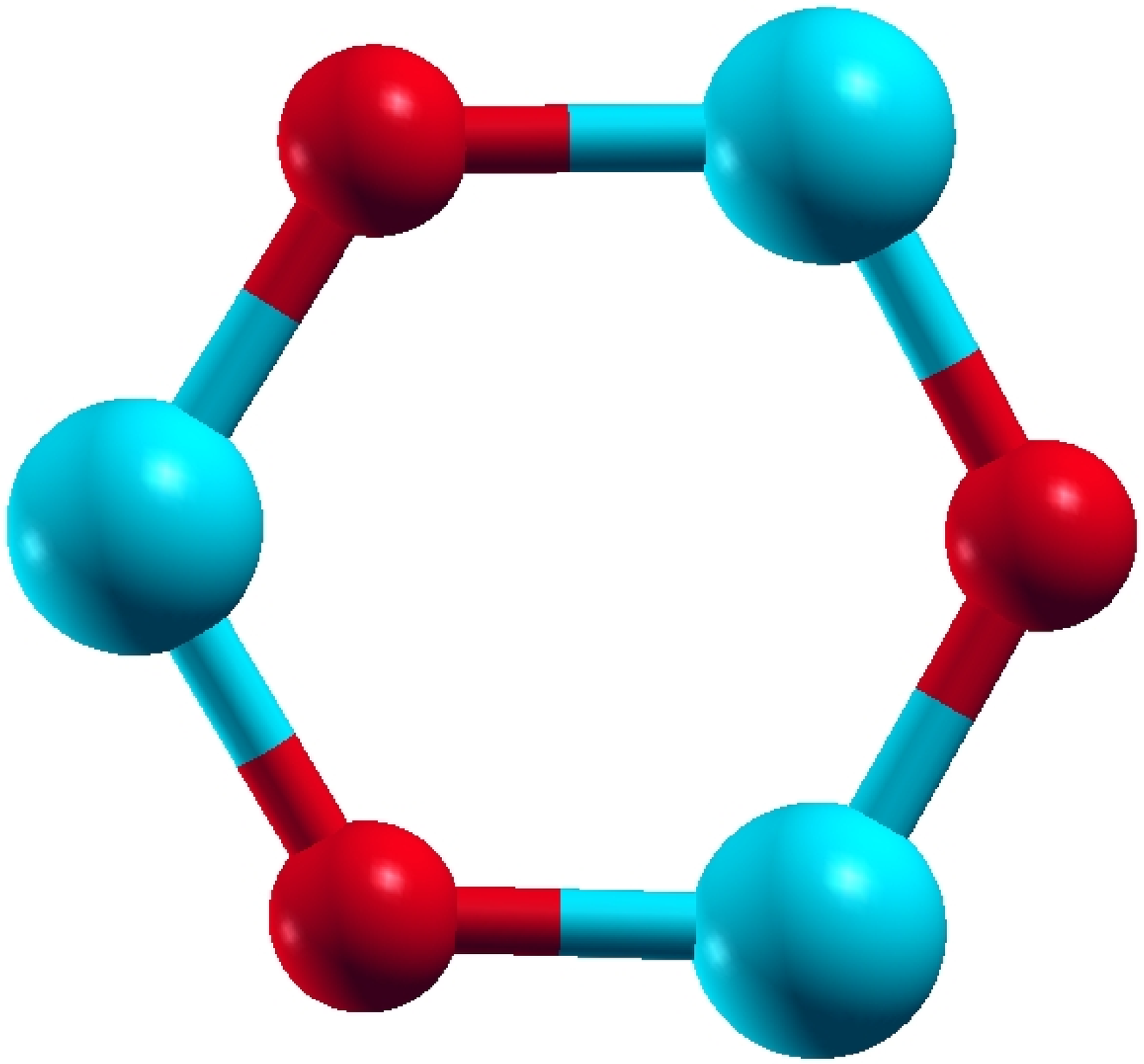} 

\hspace{0cm} 1a \hspace{5cm} 2a \hspace{4.5cm} 3a \\
\hspace{0cm}  $C_{\infty v}$ \hspace{5cm}  $D_{2h}$ \hspace{4.5cm} $D_{3h}$
\label{1-2-3LiF}
\end{figure}

\begin{figure}
\caption{ (Color online)  
Structures of (LiF)$_4$ clusters. For the notation, cf. fig. \ref{1-2-3LiF}.}
\includegraphics[width=3.5cm]{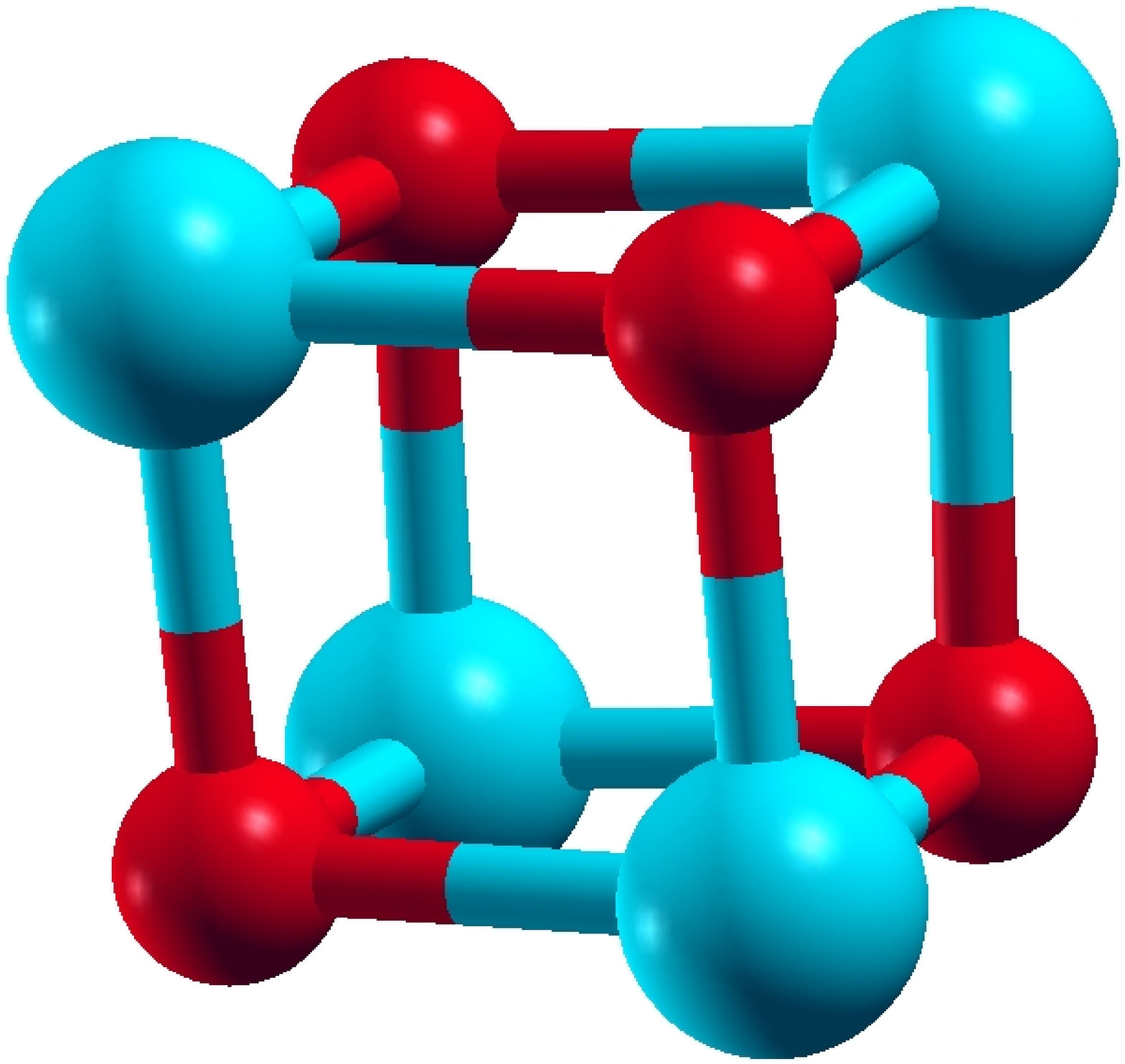} 
\includegraphics[width=5cm]{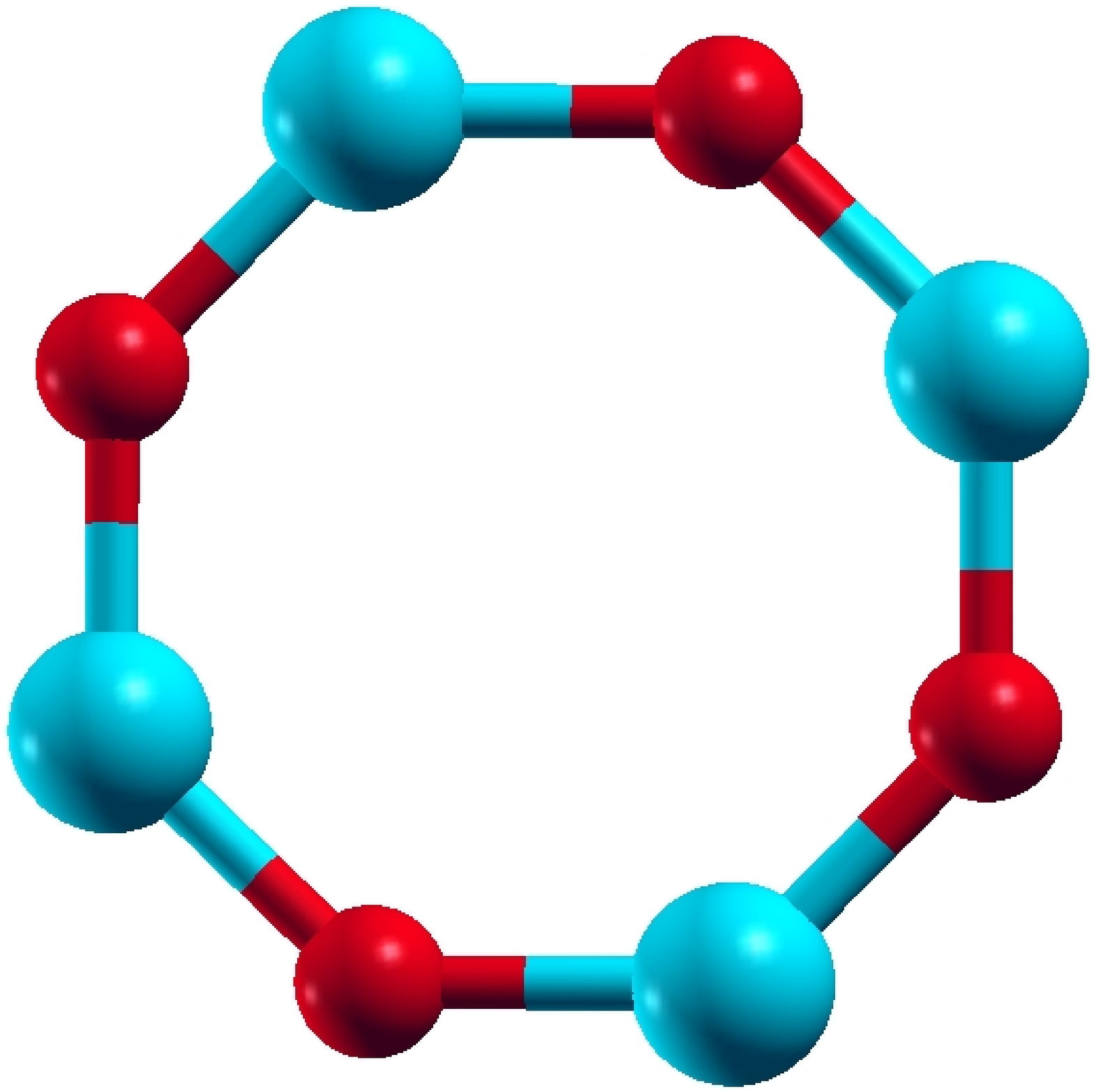}  
\includegraphics[width=5cm]{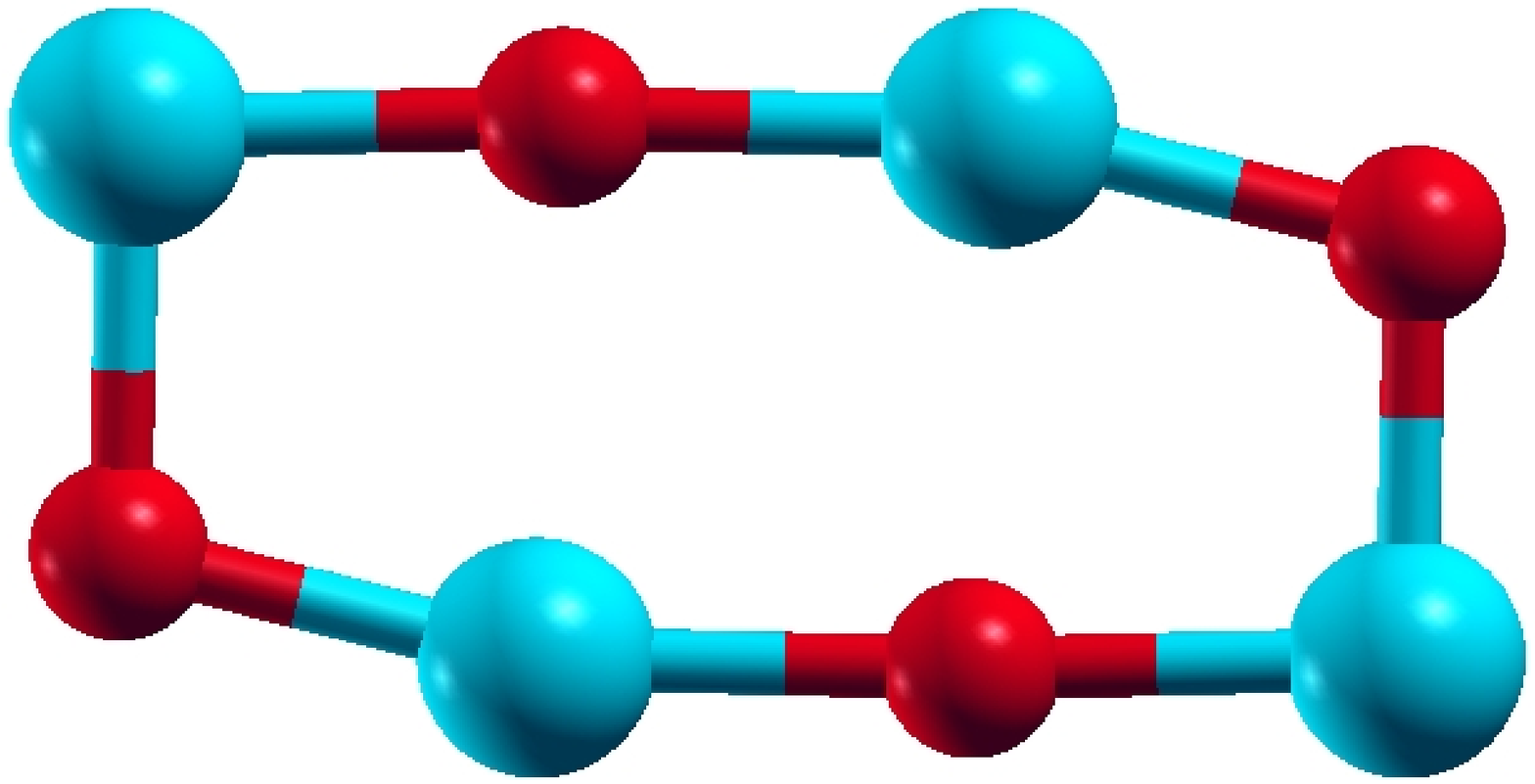} 

\hspace{0cm} 4a  \hspace{5cm} 4b  \hspace{4.5cm} 4c \\
\hspace{0cm} $T_d$ \hspace{5cm}  $D_{4h}$ \hspace{4.5cm}  $C_{2h}$
\vspace{.5cm}

\includegraphics[width=5cm]{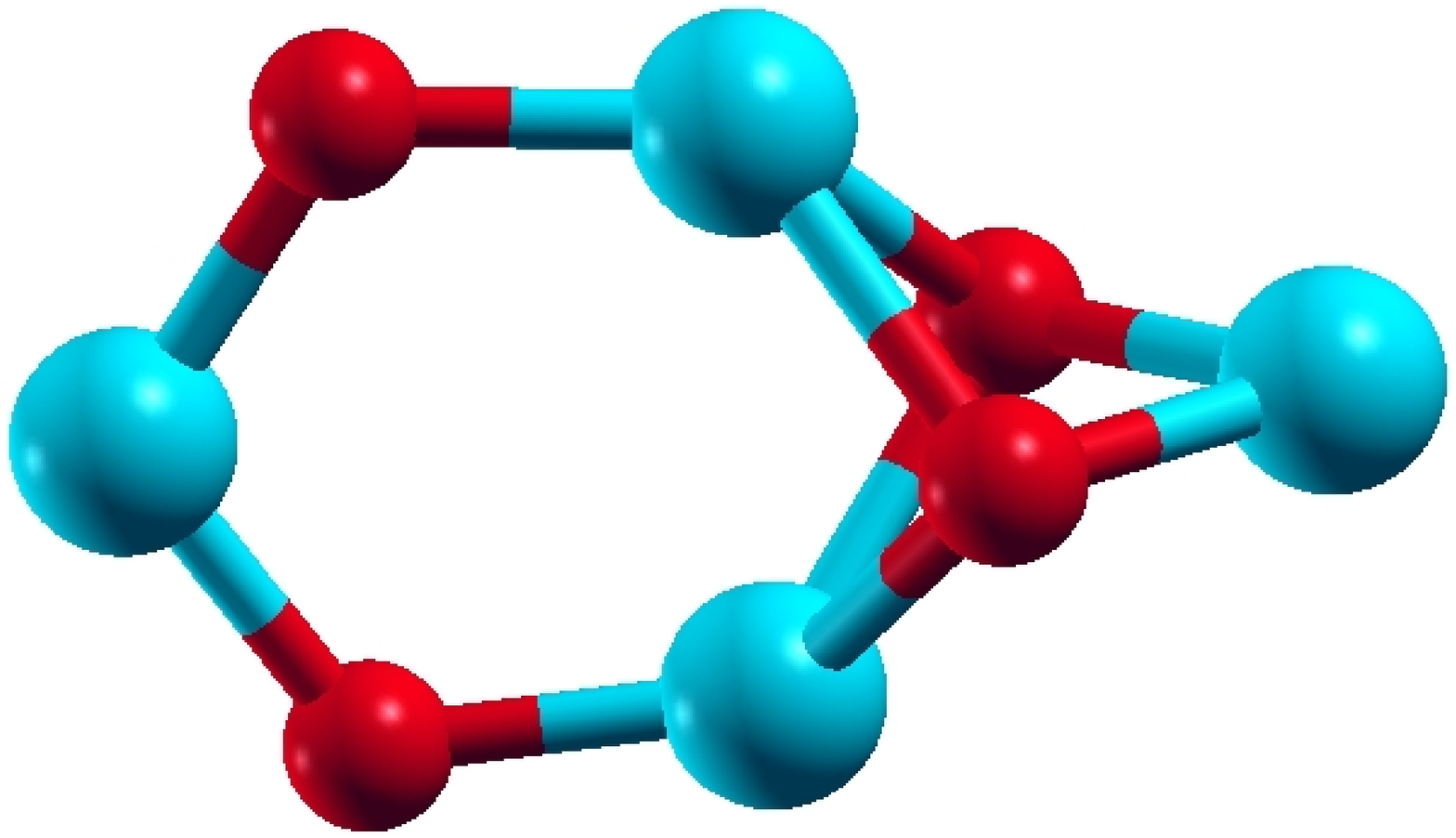} 
\includegraphics[width=5cm]{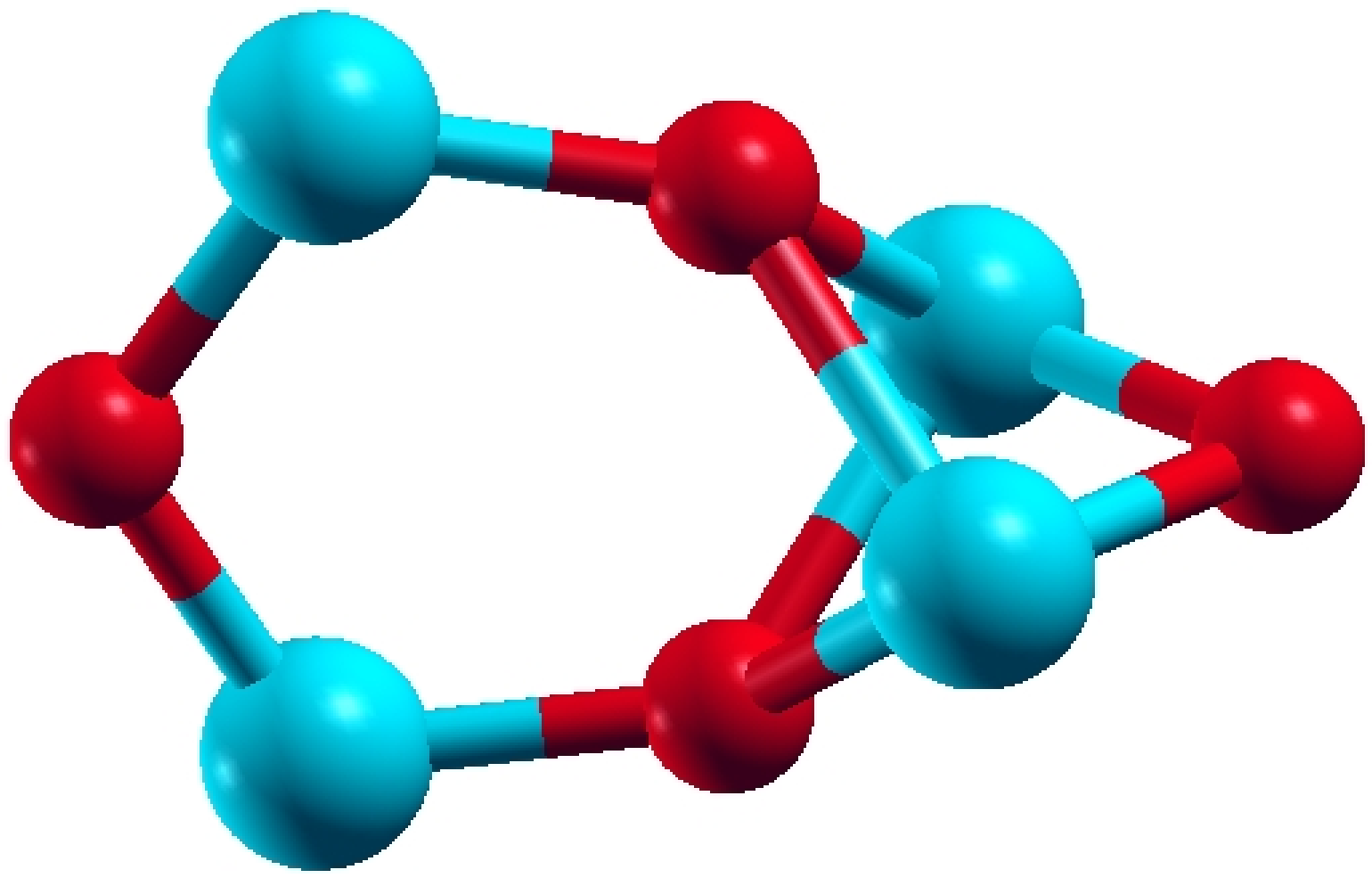}

\hspace{0cm} 4d  \hspace{5cm} 4e  \\
\hspace{0cm} $C_{2v}$ \hspace{5cm}  $C_{2v}$

\label{4LiF}
\end{figure}

\clearpage

\begin{figure}
\caption{ (Color online) 
Structures of (LiF)$_5$ clusters. For the notation, cf. fig. \ref{1-2-3LiF}.}
\includegraphics[width=4cm]{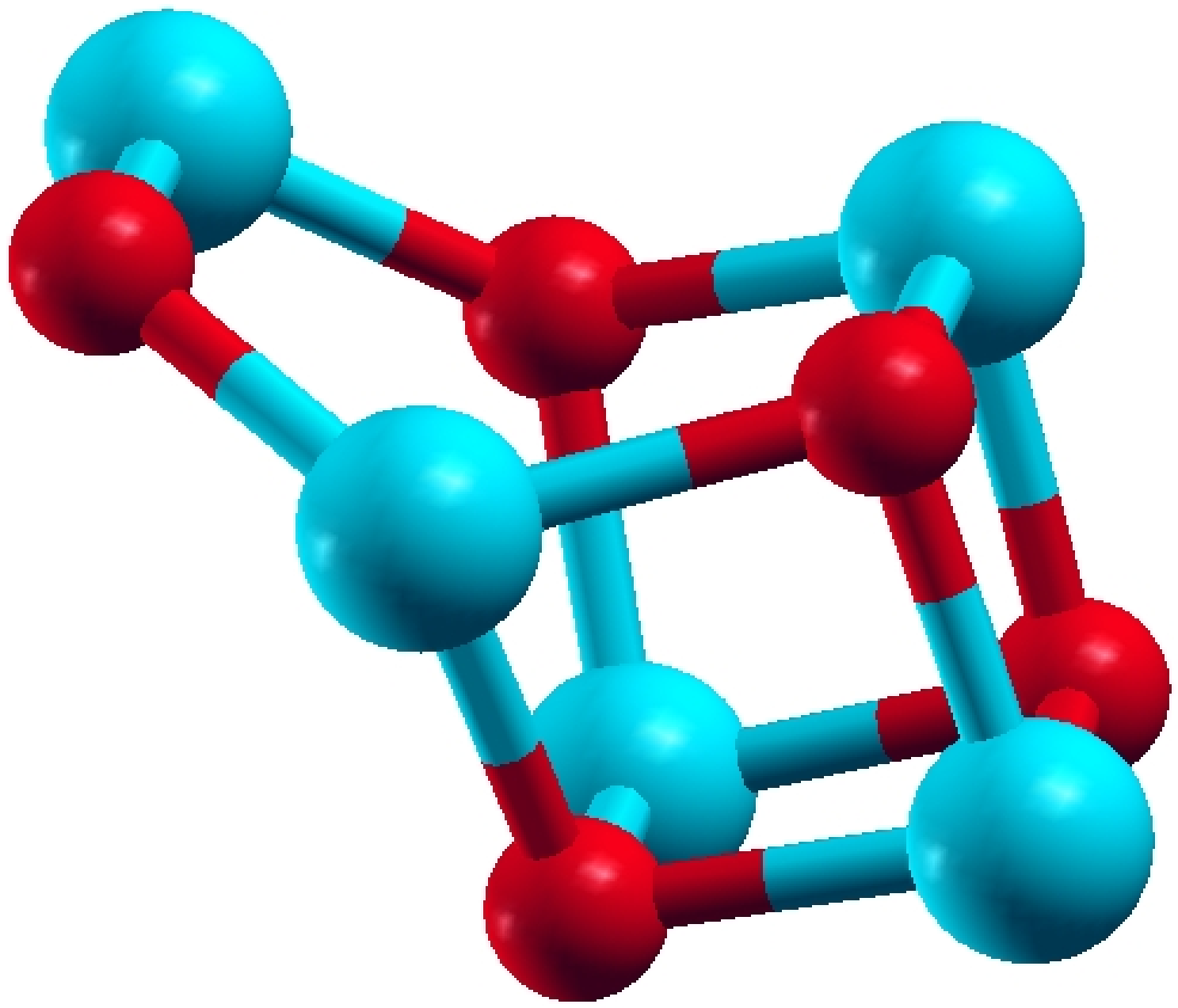} 
\includegraphics[width=5cm]{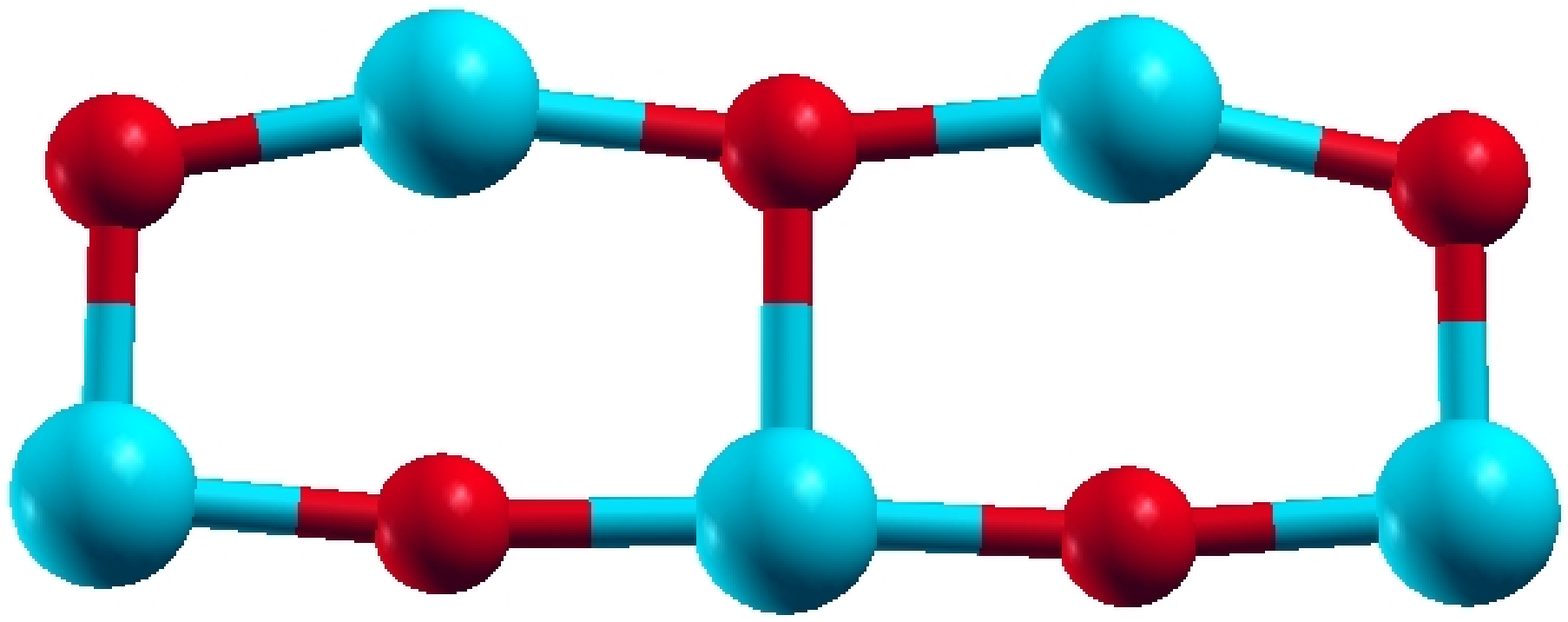} 
\includegraphics[width=5cm]{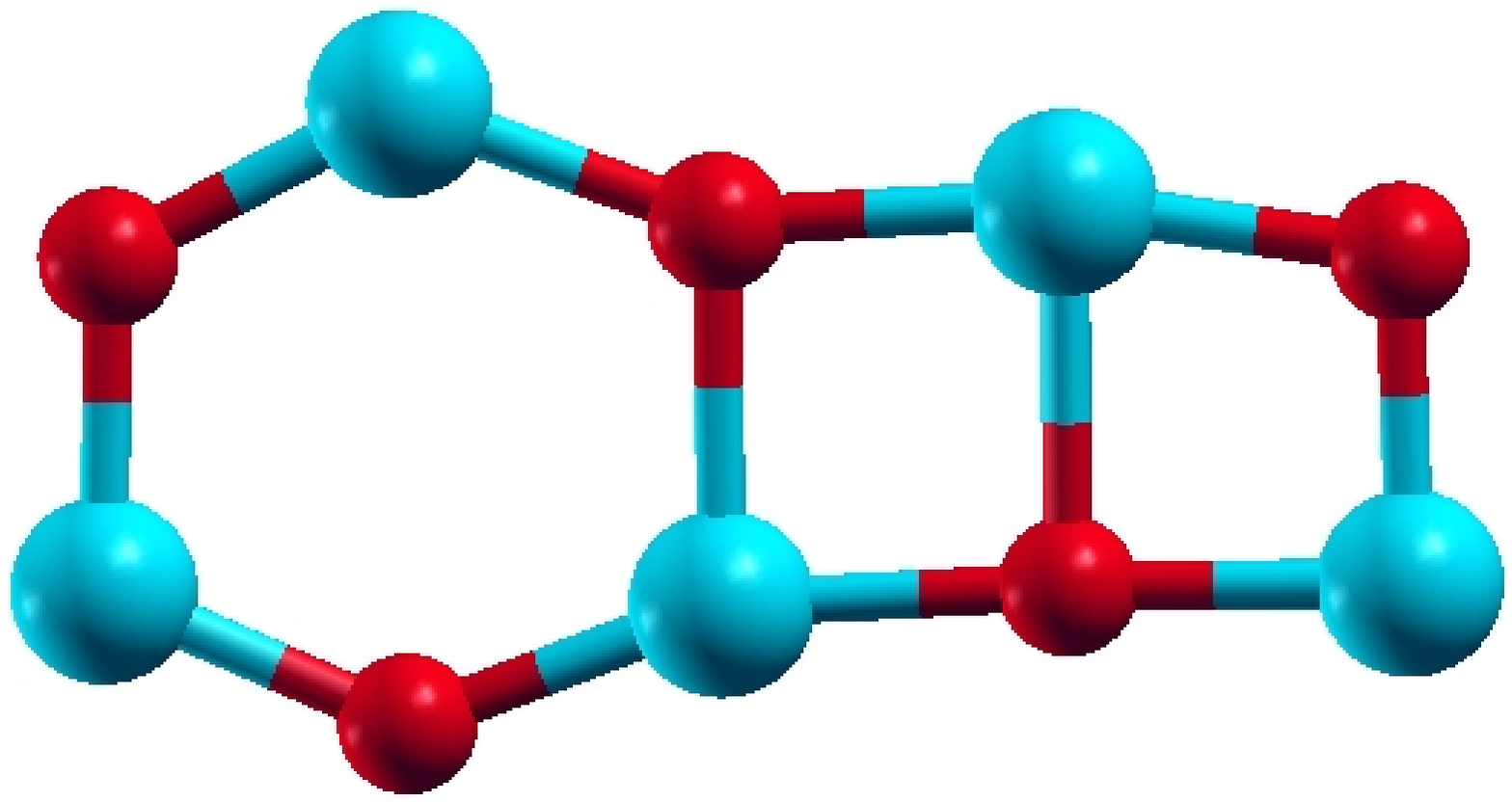} 

\hspace{0cm} 5a  \hspace{5cm} 5b \hspace{4.5cm} 5c \\
\hspace{0cm} $C_s$ \hspace{5cm}  $C_{2v}$ \hspace{4.5cm} $C_s$
\vspace{.5cm}

\includegraphics[width=5cm]{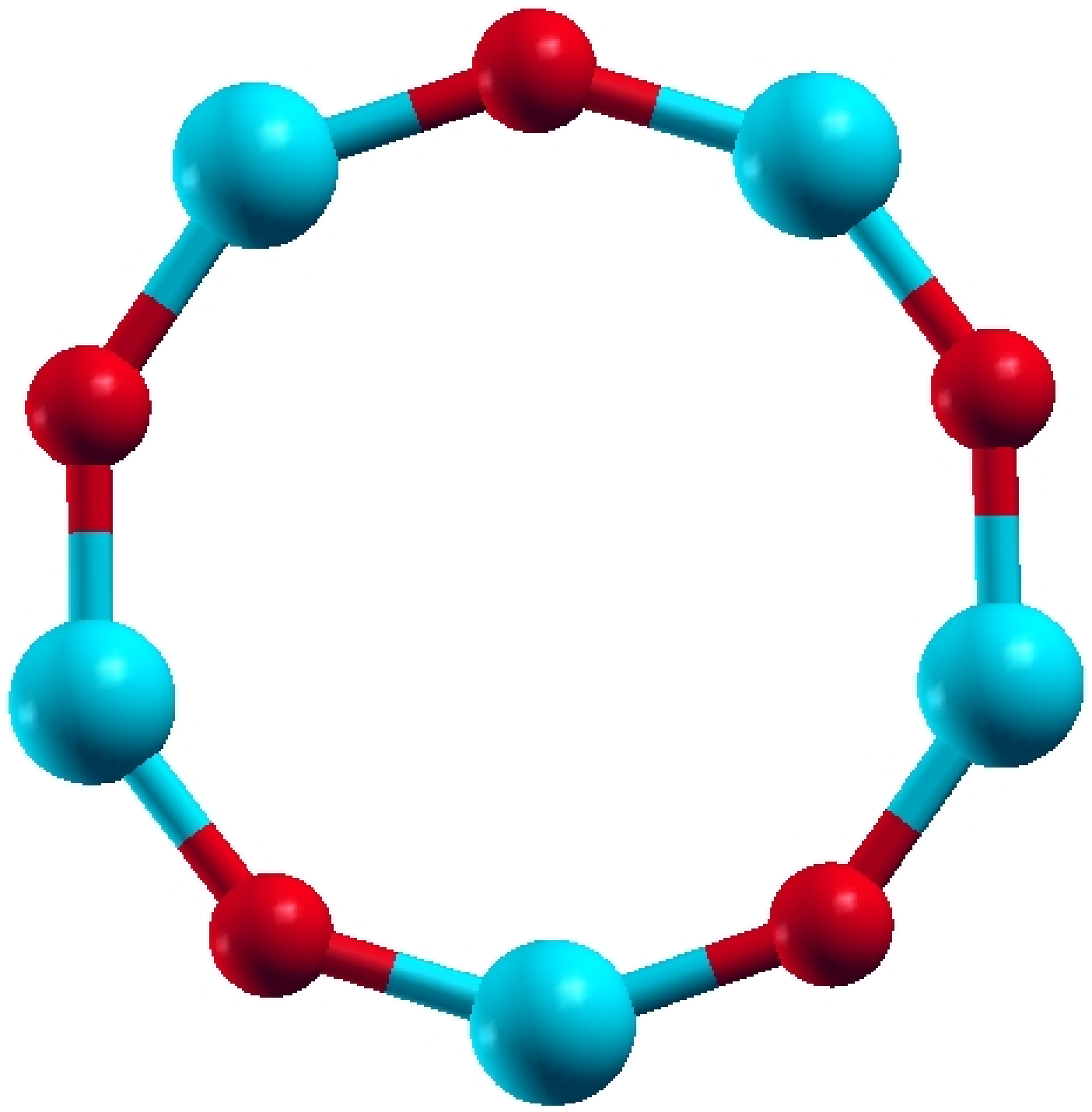} 
\includegraphics[width=5cm]{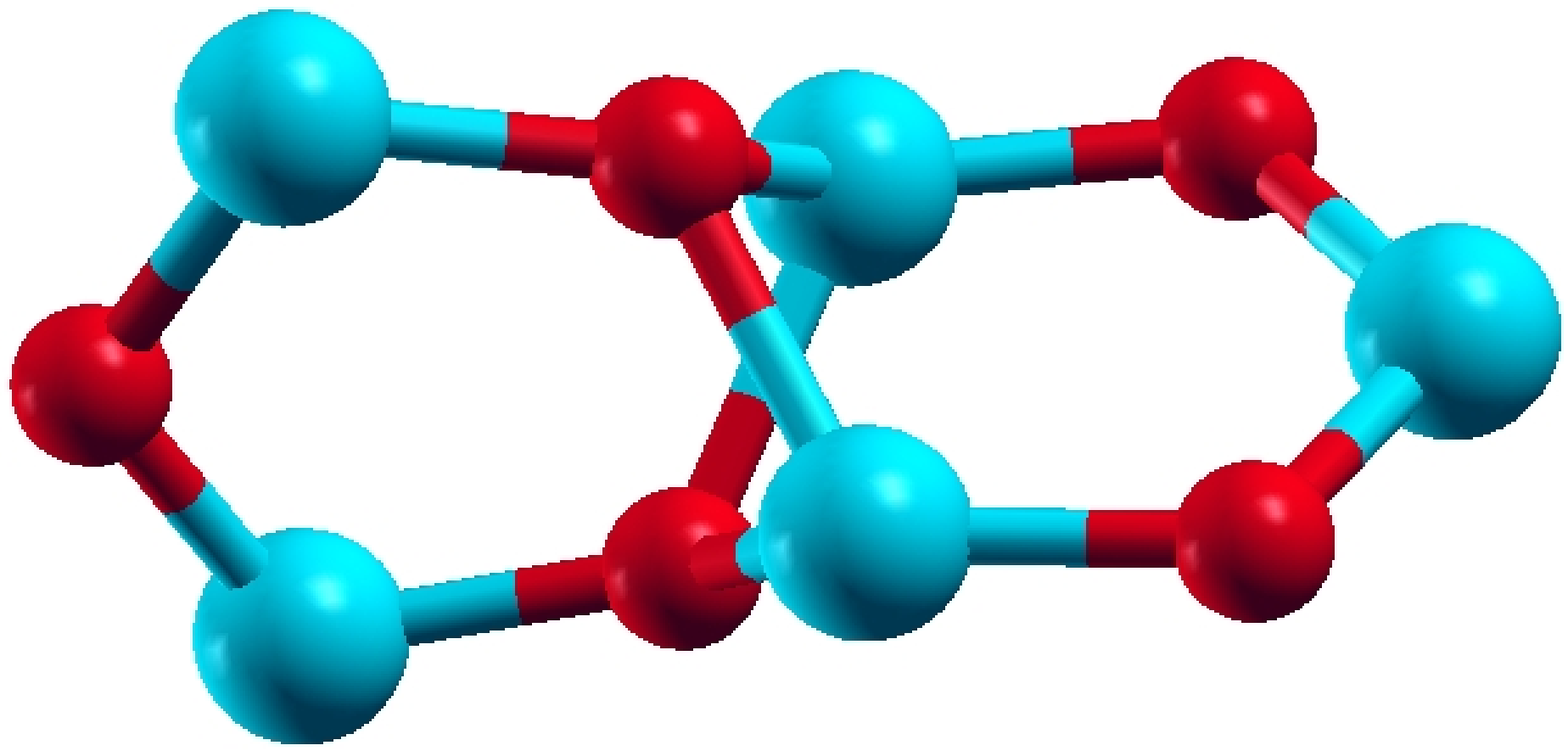} 
\includegraphics[width=5cm]{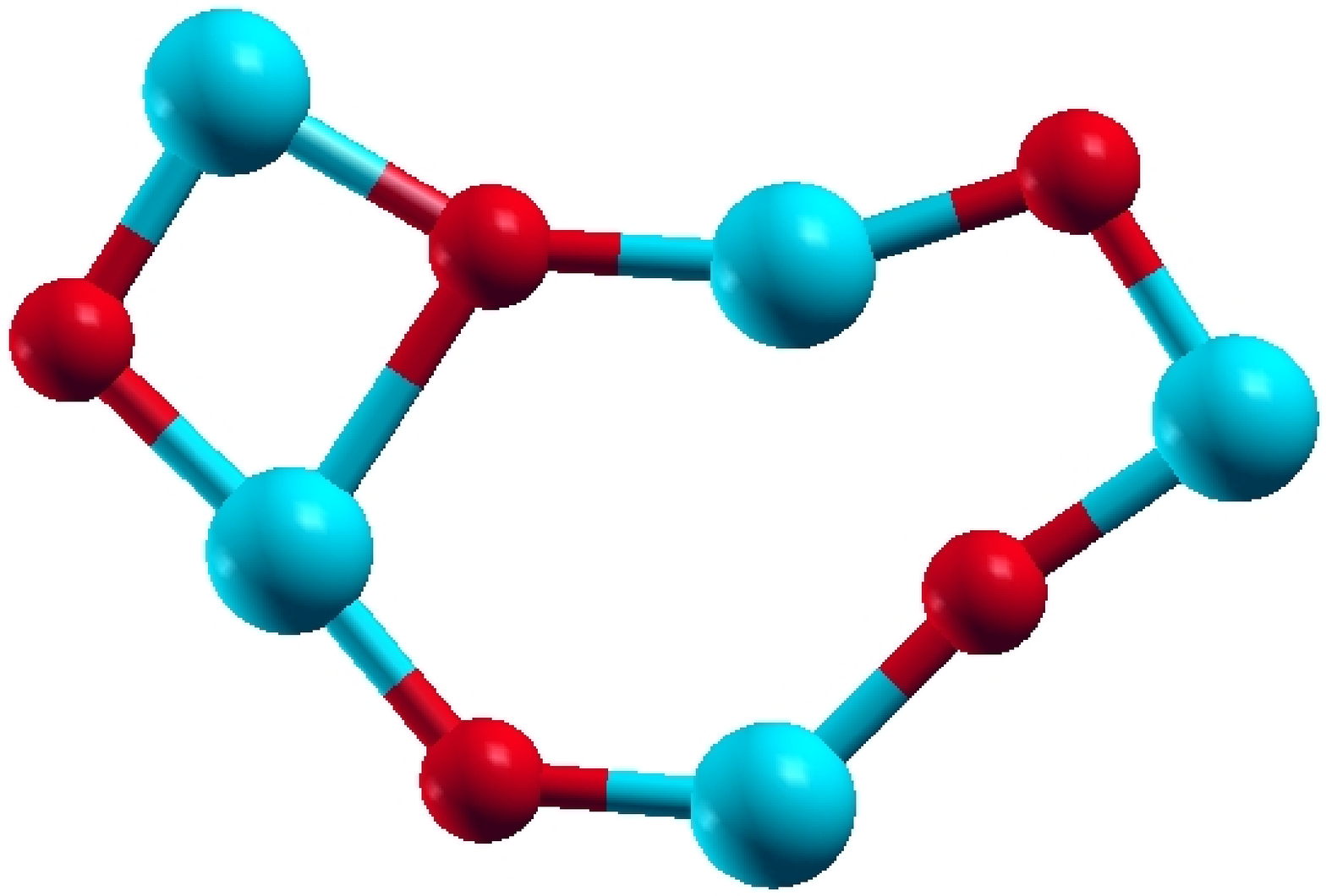} 

\hspace{0cm} 5d \hspace{5cm} 5e \hspace{4.5cm} 5f \\
\hspace{0cm} $D_{5h}$ \hspace{5cm} $C_{2v}$ \hspace{4.5cm} $C_s$
\vspace{.5cm}

\includegraphics[width=5cm]{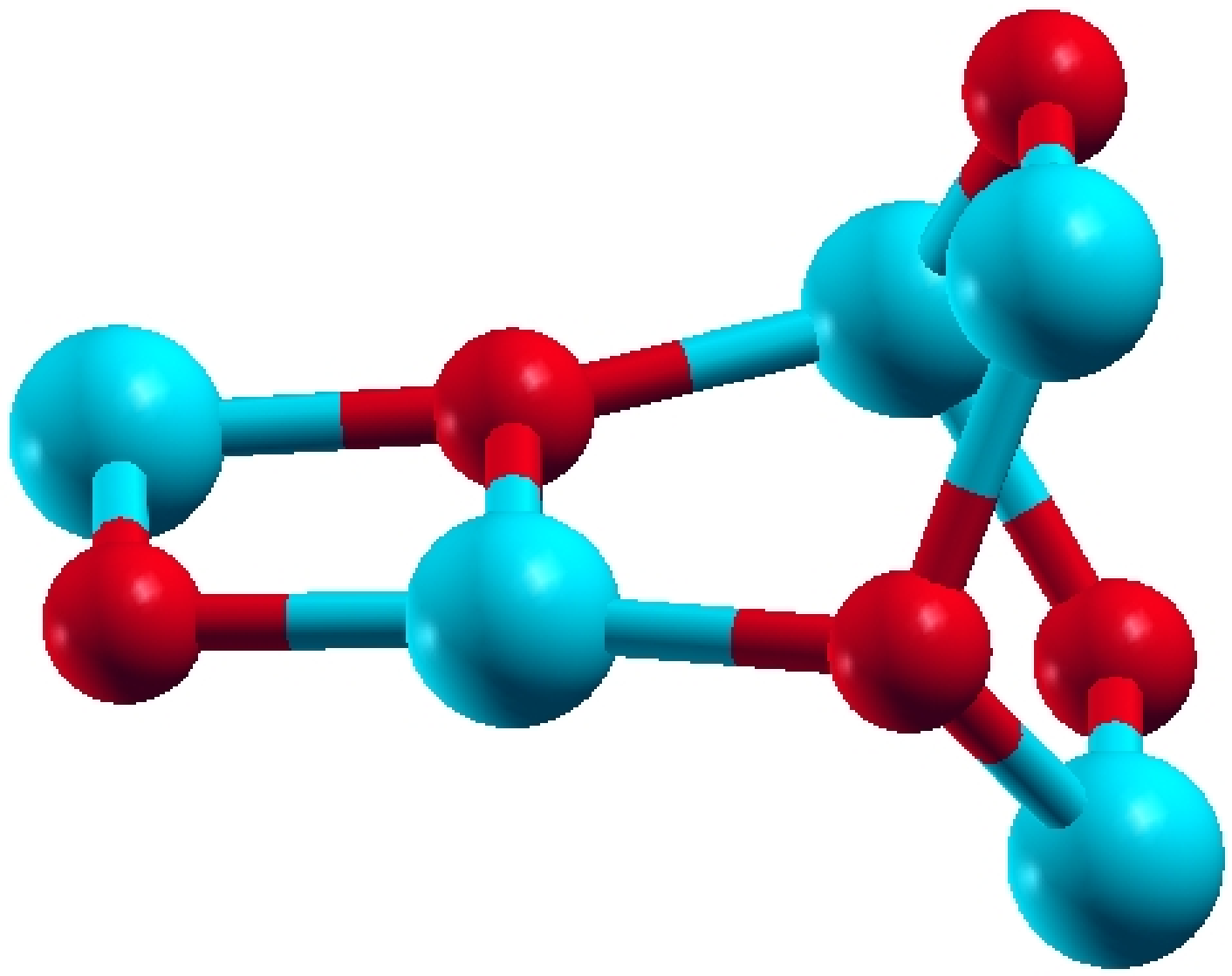} 

 5g \\
 $C_s$

\label{5LiF} 
\end{figure}

\clearpage

\begin{figure}
\caption{(Color online)  
Structures of (LiF)$_6$ clusters. For the notation, cf. fig. \ref{1-2-3LiF}.}
\includegraphics[width=4.cm]{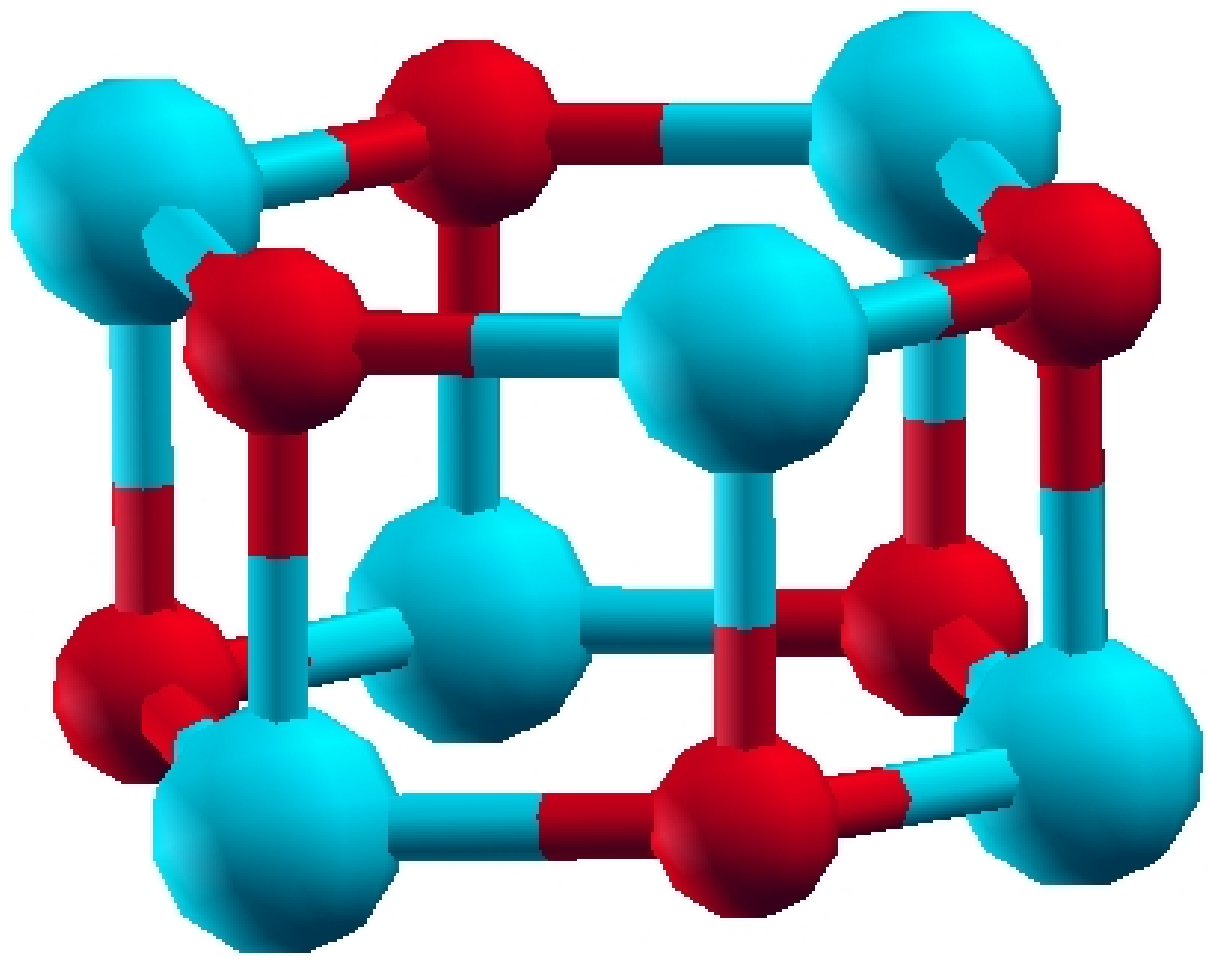} 
\includegraphics[width=4.cm]{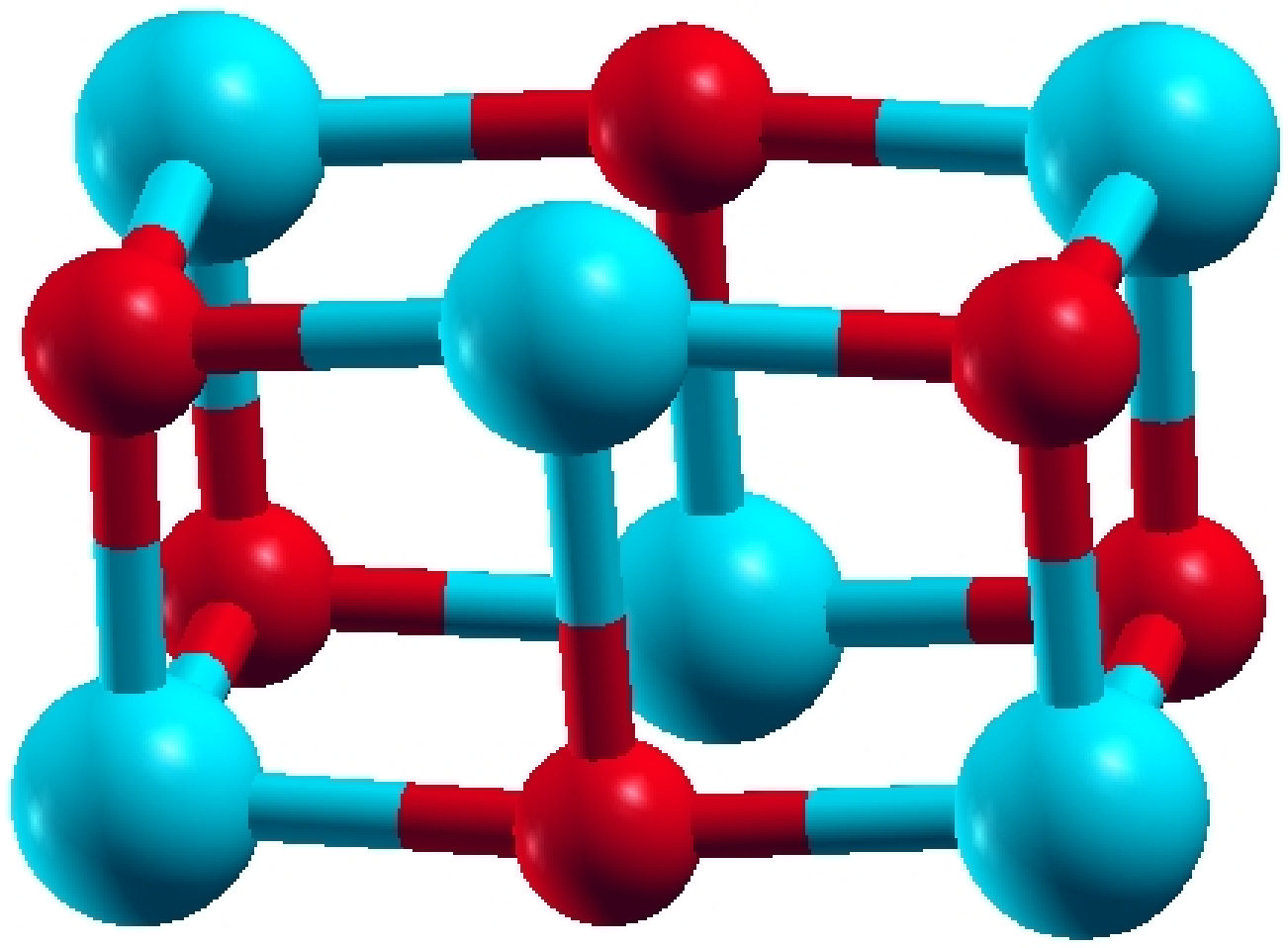} 
\includegraphics[width=4.cm]{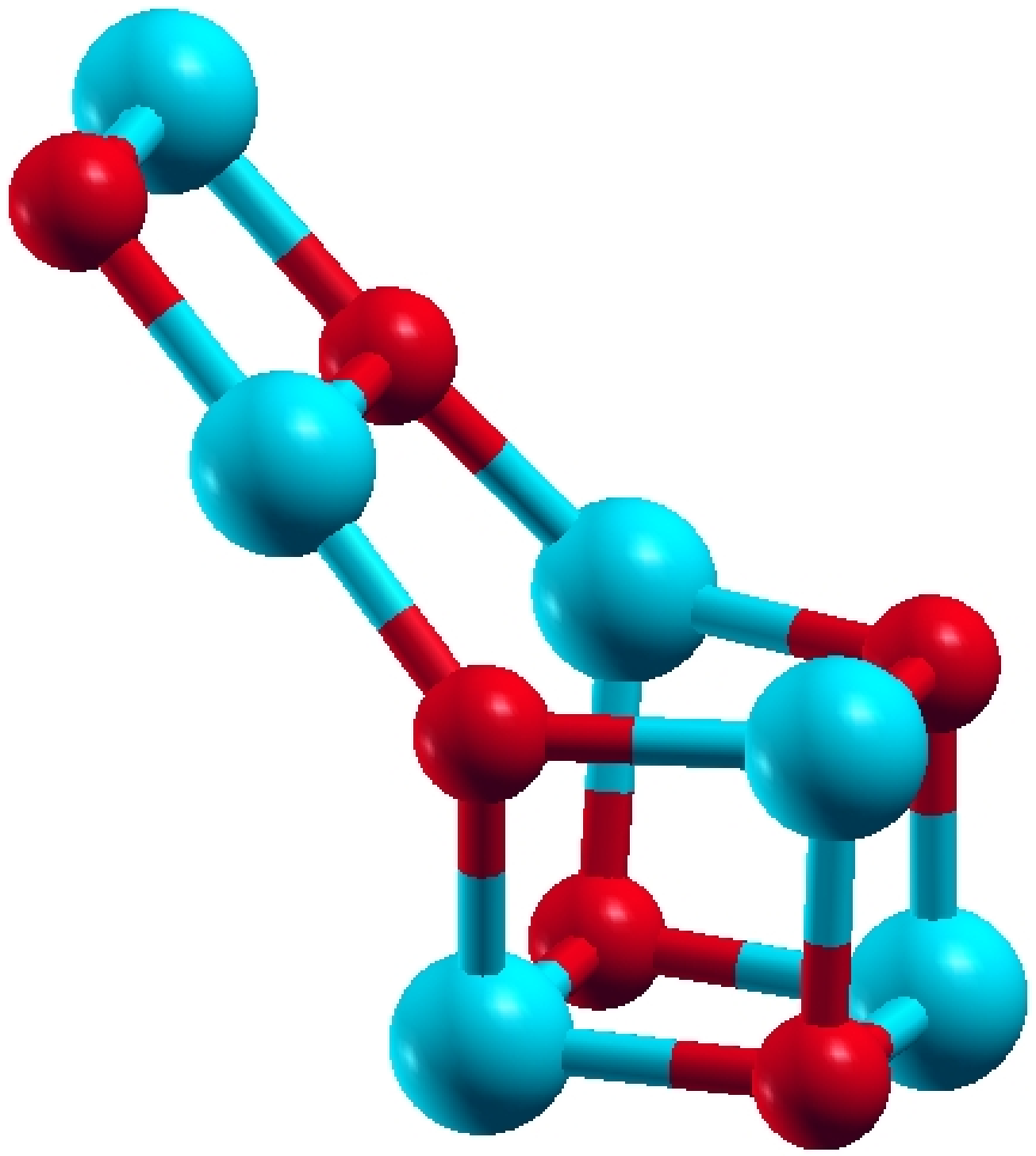}

\hspace{0cm} 6a \hspace{5cm} 6b \hspace{4.5cm} 6c \\
\hspace{0cm} $D_{3d}$ \hspace{5cm} $D_{2h}$ \hspace{4.5cm} $C_s$
\vspace{.5cm}

\includegraphics[width=5cm]{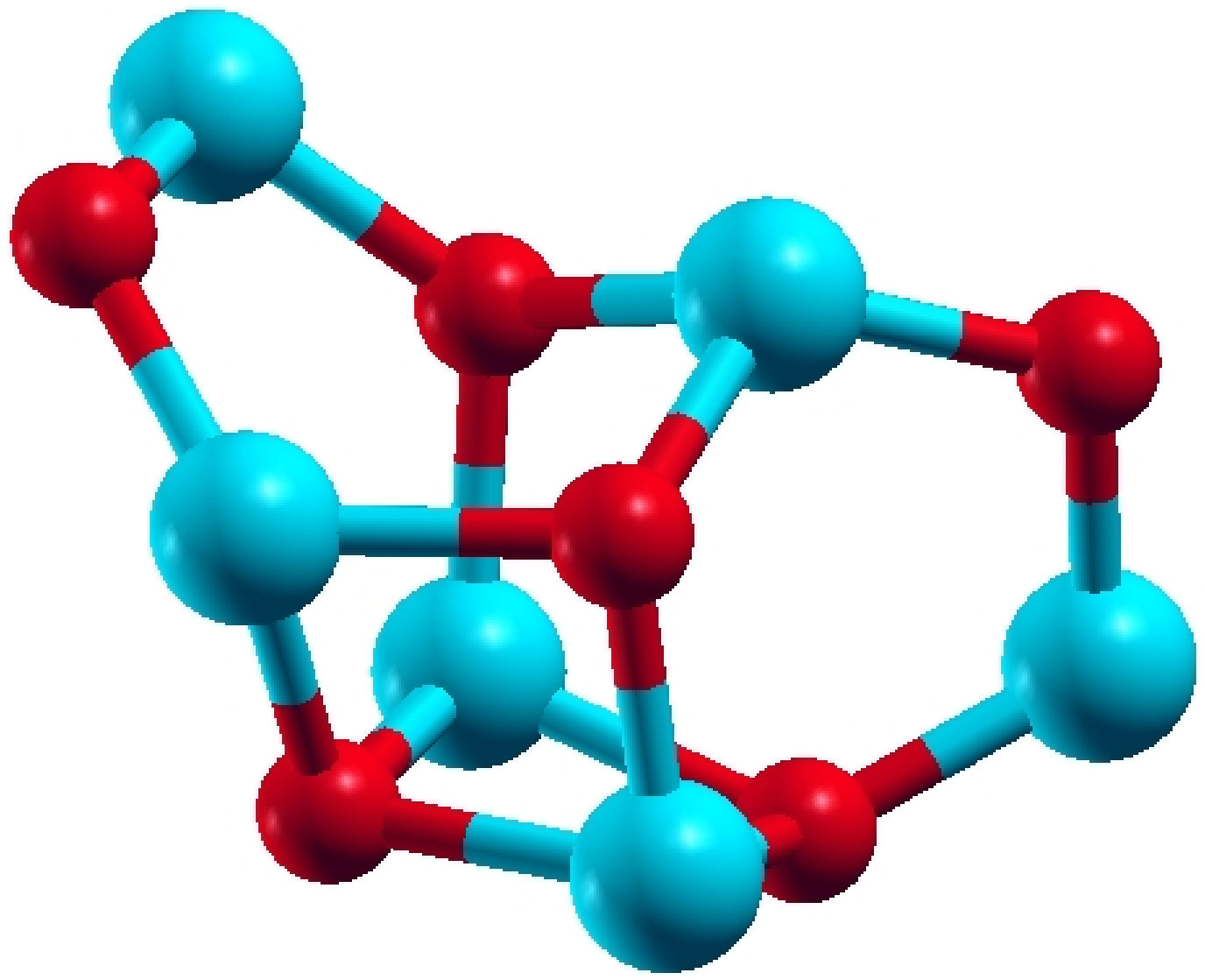}
\includegraphics[width=5cm]{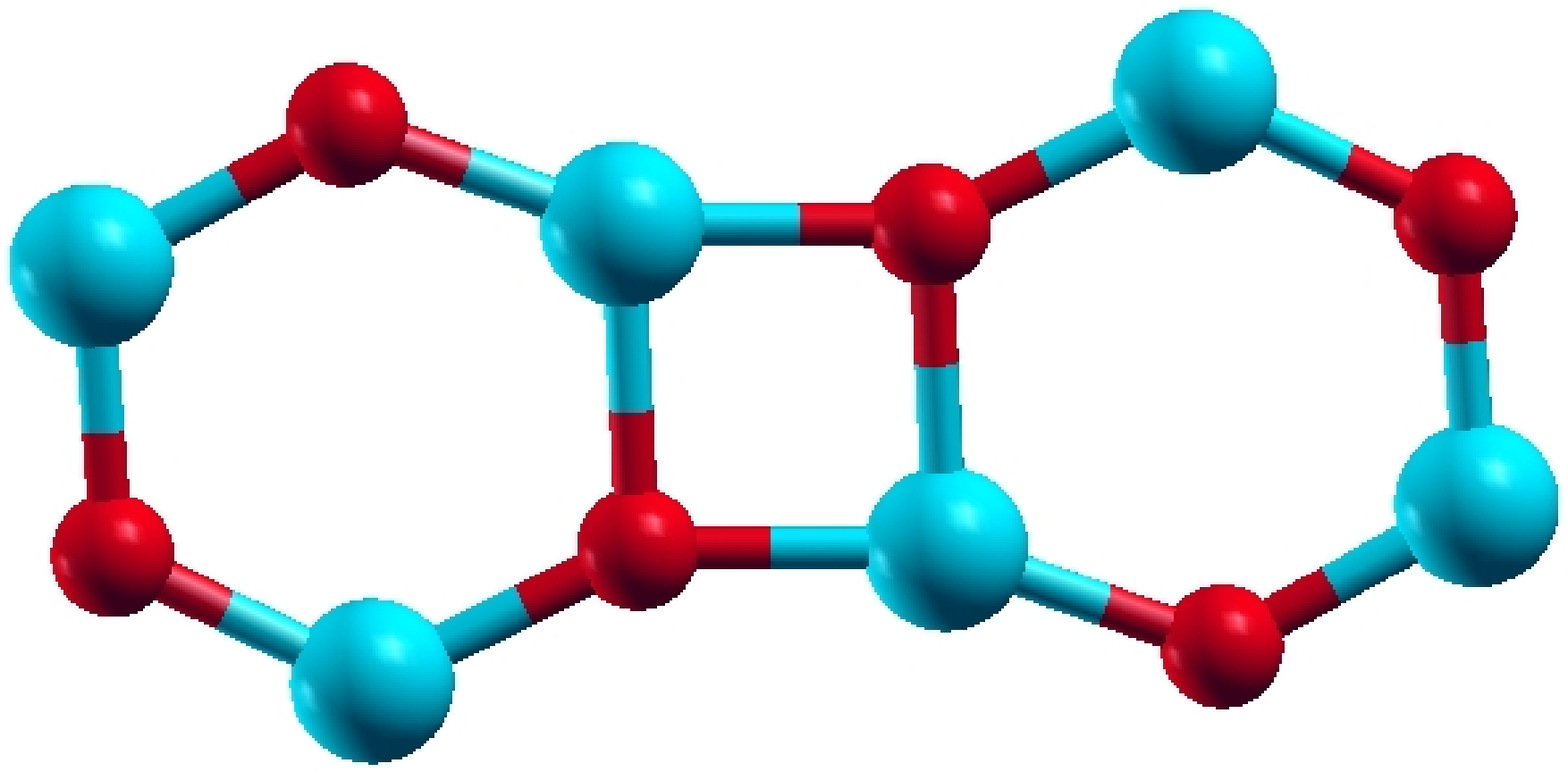}
\includegraphics[width=5cm]{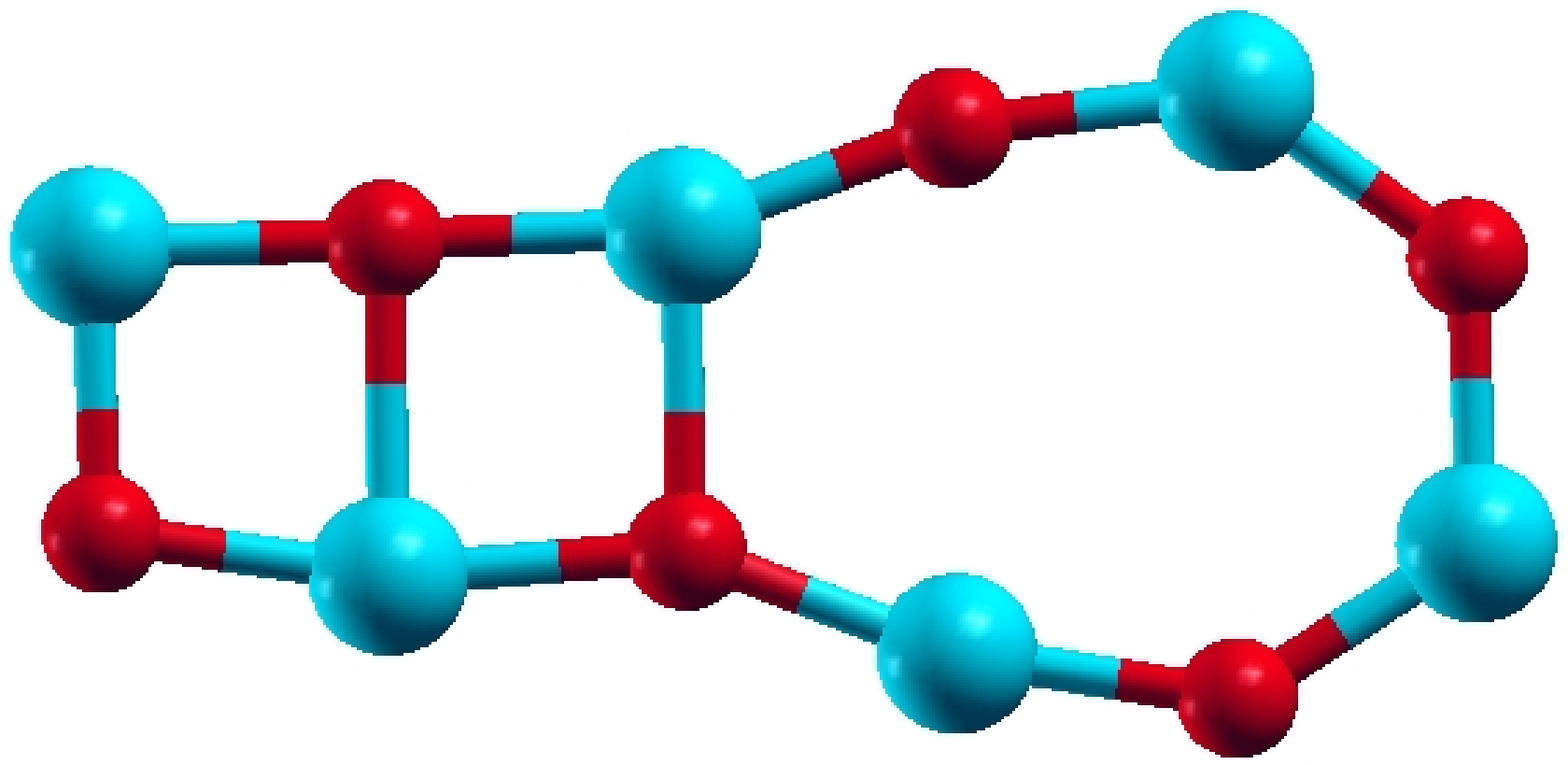}

\hspace{-2.5cm}   6d (other similar structures
possible)\hspace{2.5cm} 6e \hspace{4.5cm} 6f \\
\hspace{0cm} $C_1$ \hspace{5cm} $C_{2h}$ \hspace{4.5cm} $C_s$
\vspace{.5cm}

\includegraphics[width=5cm]{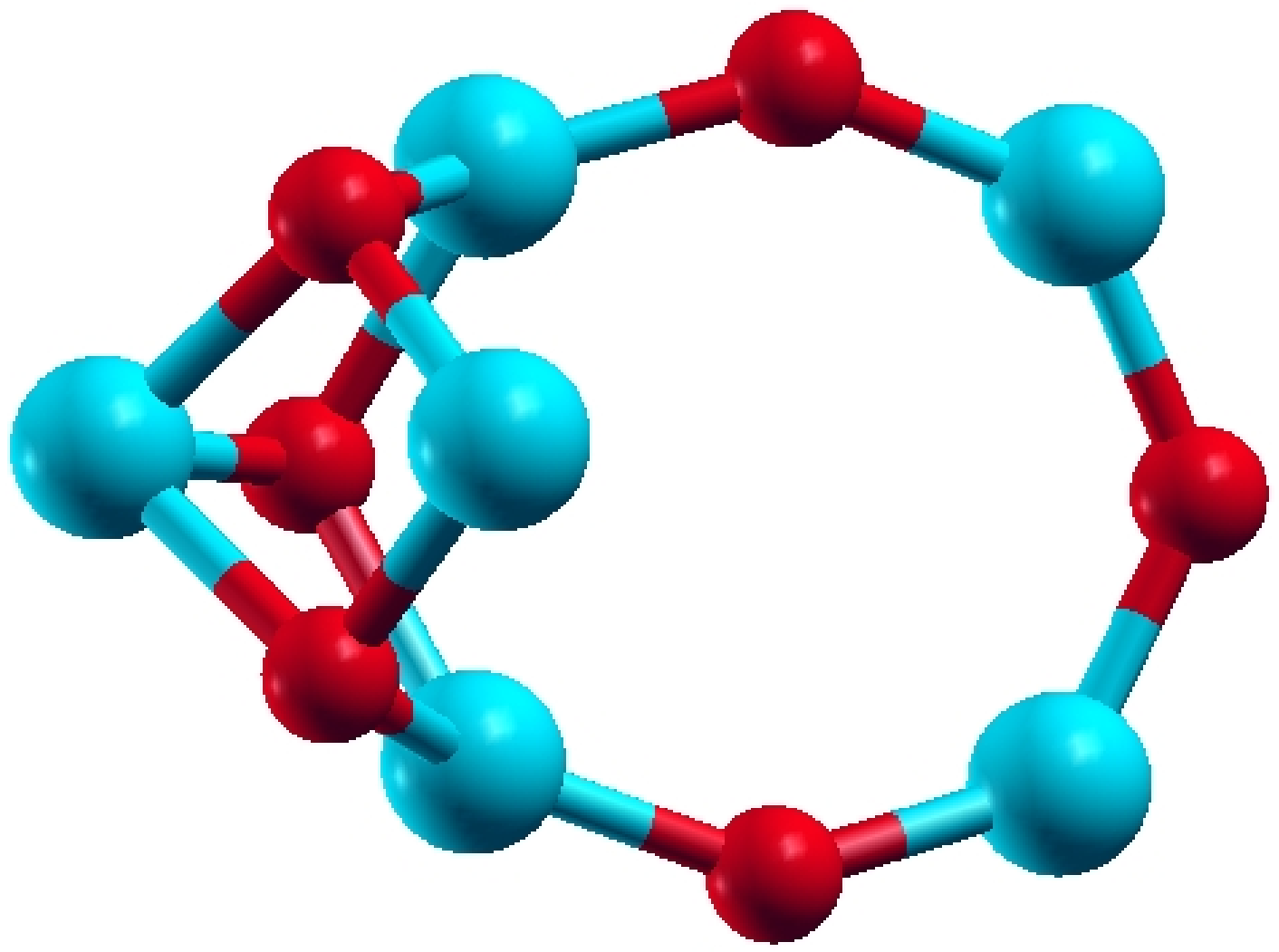}
\includegraphics[width=5cm]{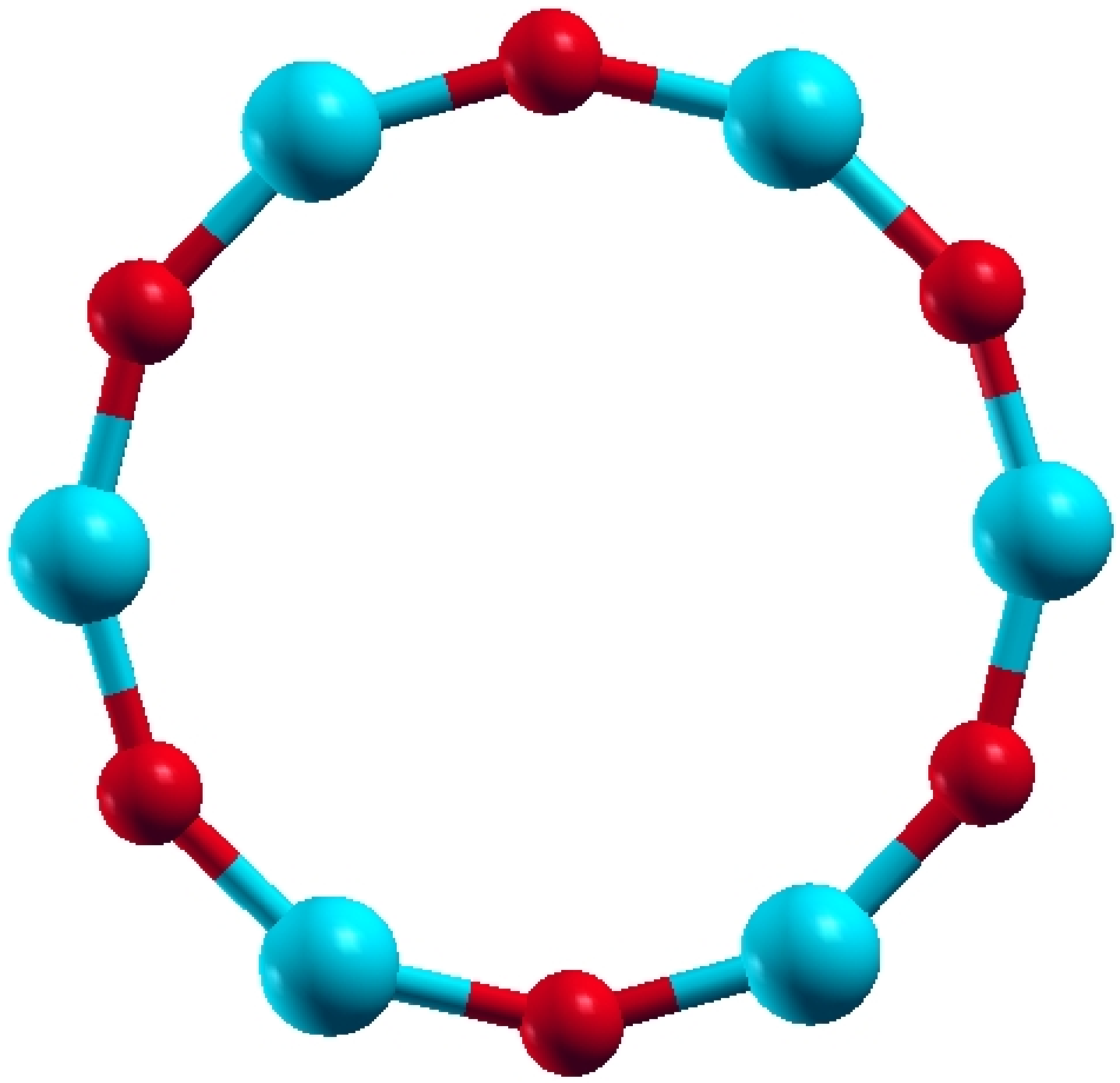} 

\hspace{0cm} 6g \hspace{5cm} 6h  \\
\hspace{0cm}   $C_s$ \hspace{5cm} $D_{6h}$

\label{6LiF}
\end{figure}

\clearpage

\begin{figure}
\caption{(Color online) 
Structures of (LiF)$_7$ clusters. For the notation, cf. fig. \ref{1-2-3LiF}.}
\includegraphics[width=5cm]{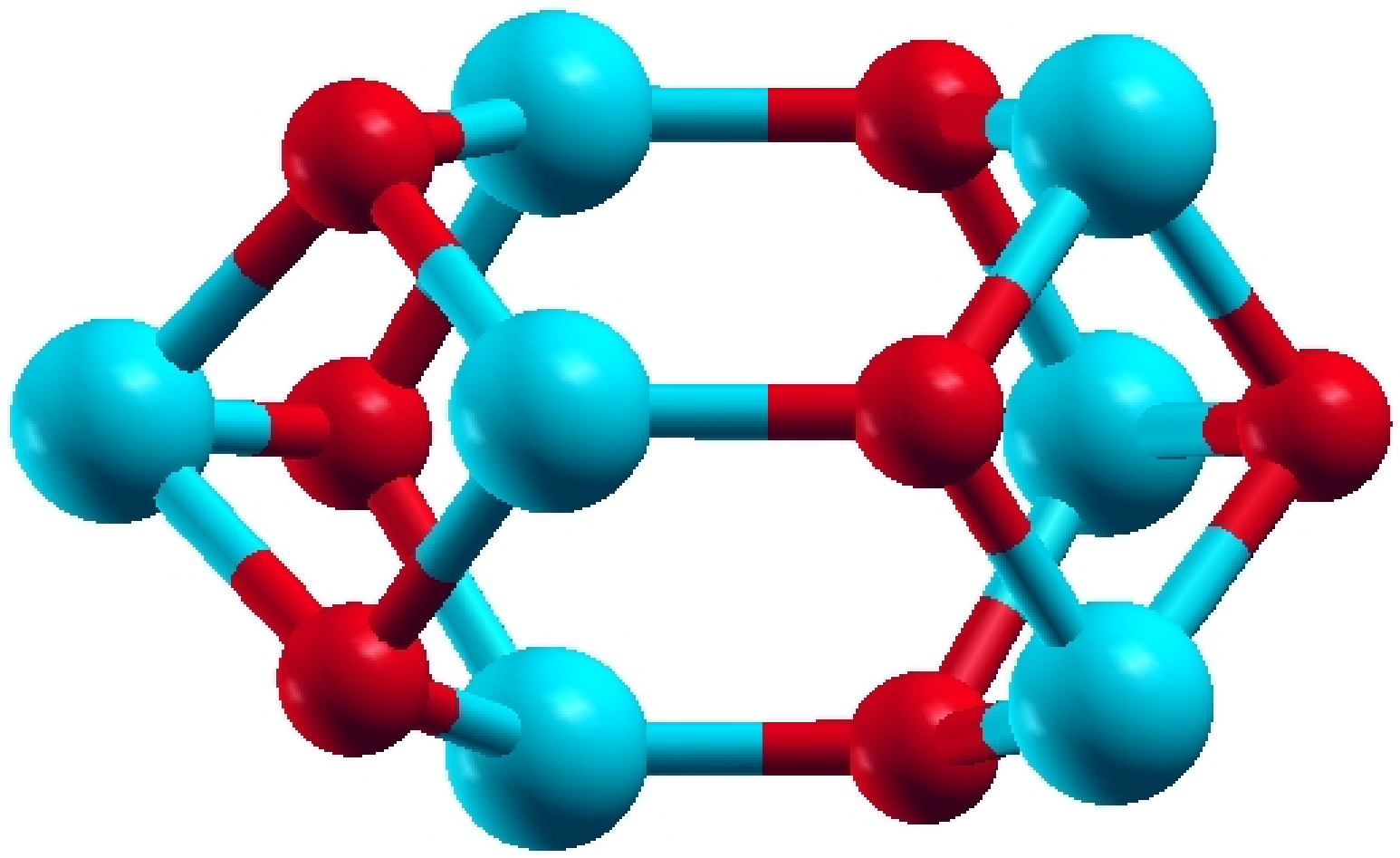} 
\includegraphics[width=5cm]{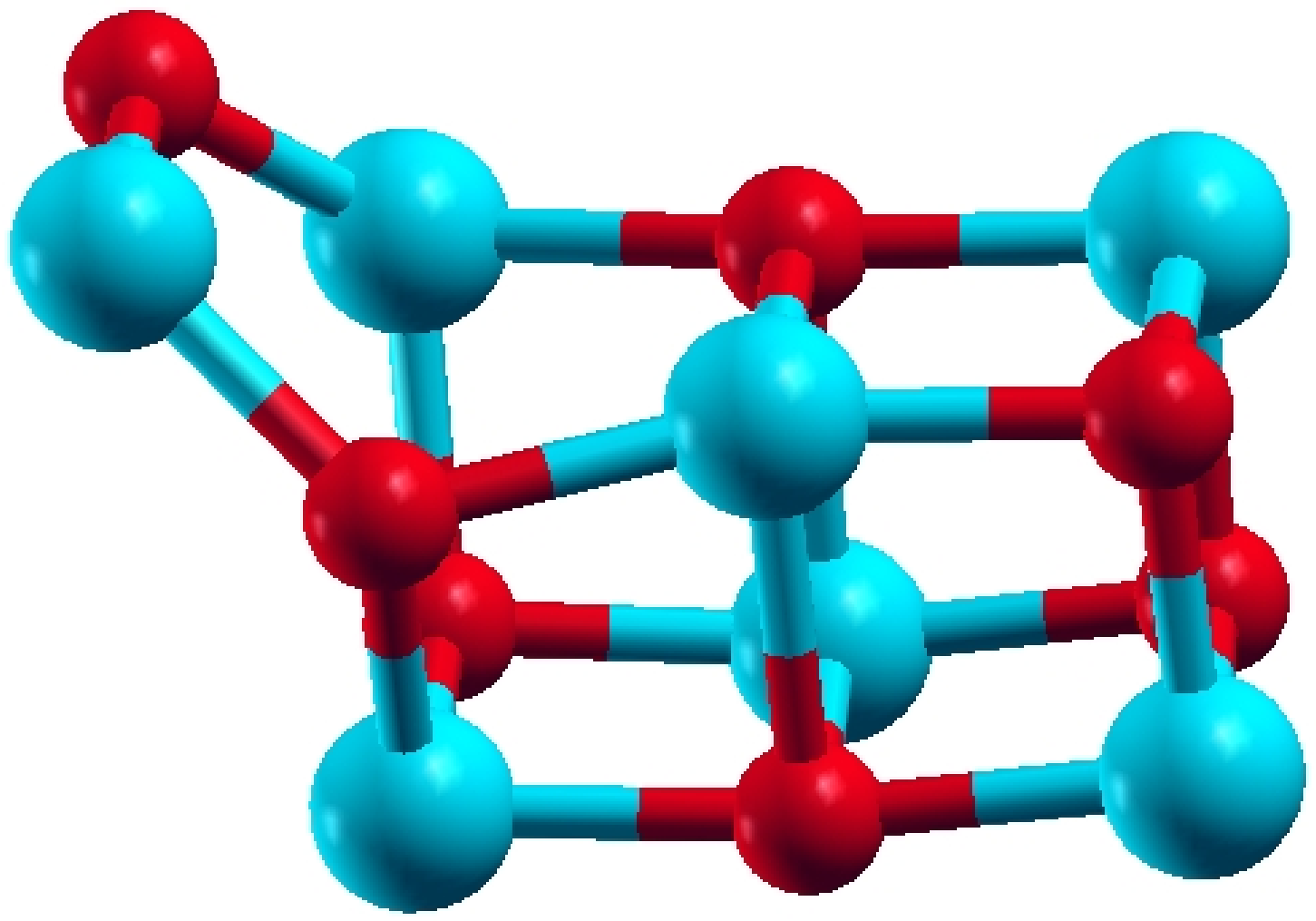} 
\includegraphics[width=5cm]{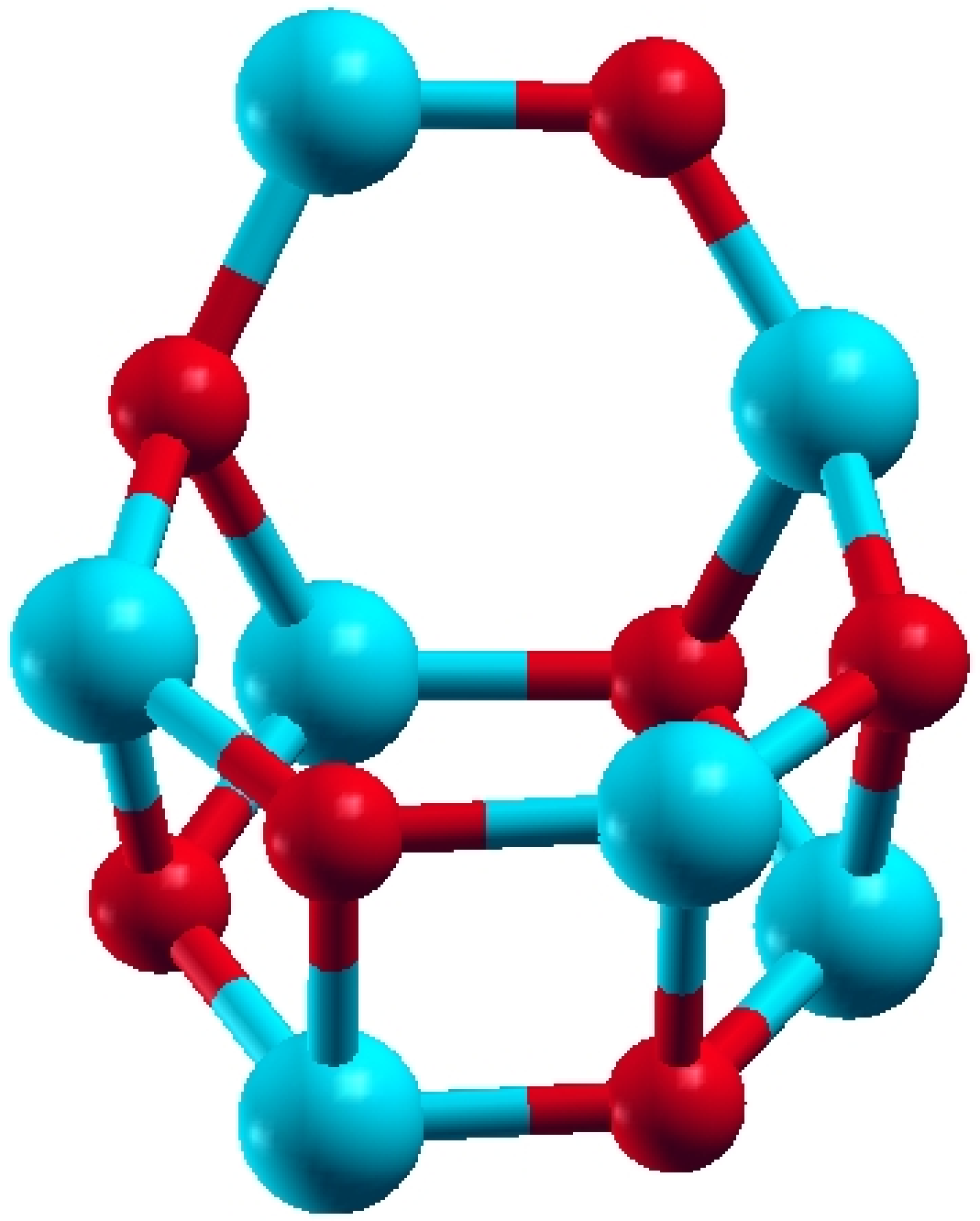}

\hspace{0cm} 7a \hspace{5cm} 7b \hspace{4.5cm} 7c \\
\hspace{0cm} $C_{3v}$ \hspace{5cm} $C_1$ \hspace{4.5cm} $C_1$
\vspace{.5cm}

\includegraphics[width=5cm]{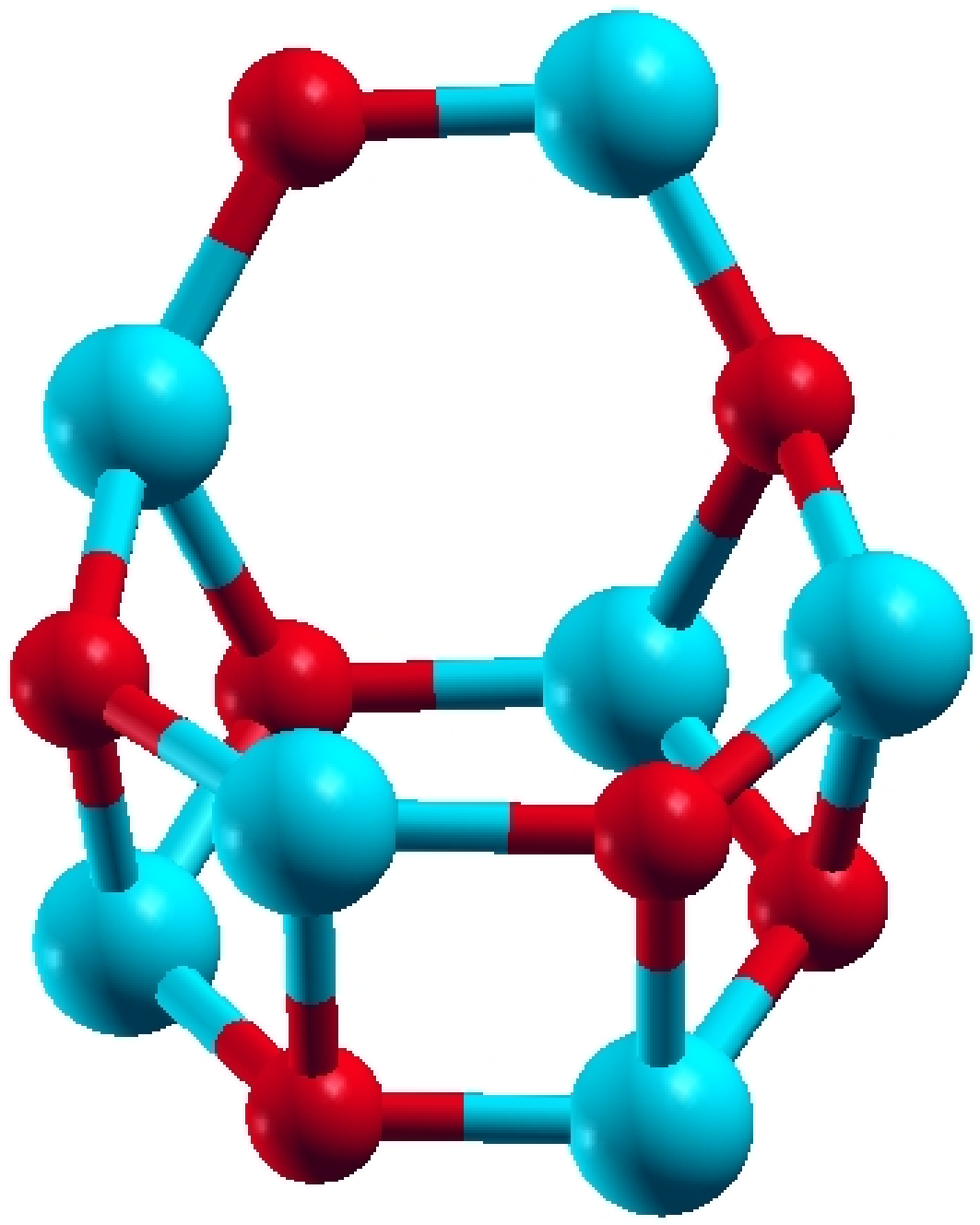} 
\includegraphics[width=5cm]{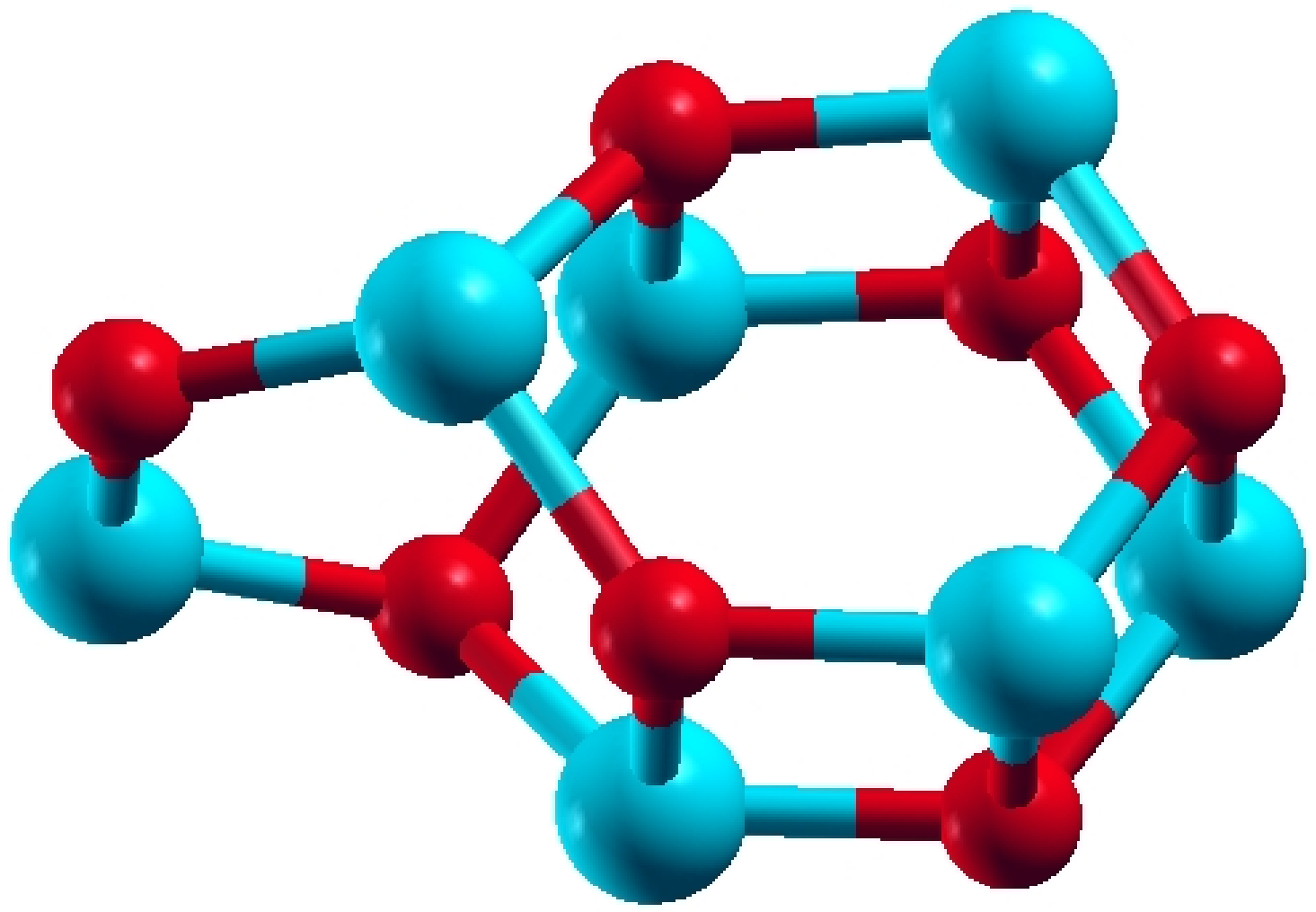} 
\includegraphics[width=5cm]{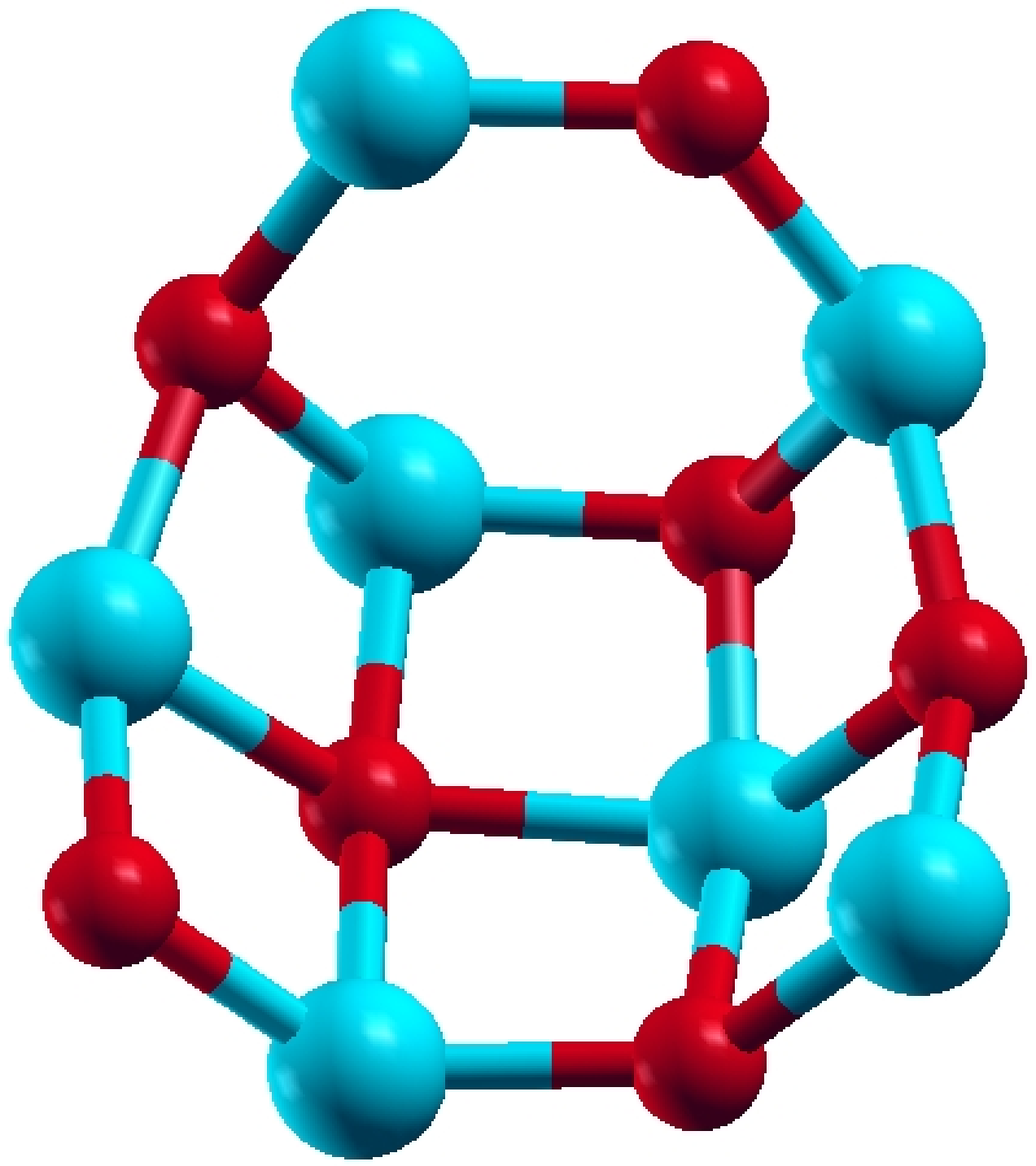} 

\hspace{0cm} 7d \hspace{5cm} 7e \hspace{4.5cm} 7f \\
\hspace{0cm} $C_1$ \hspace{5cm} $C_s$ \hspace{4.5cm} $C_1$
\vspace{.5cm}

\includegraphics[width=5cm]{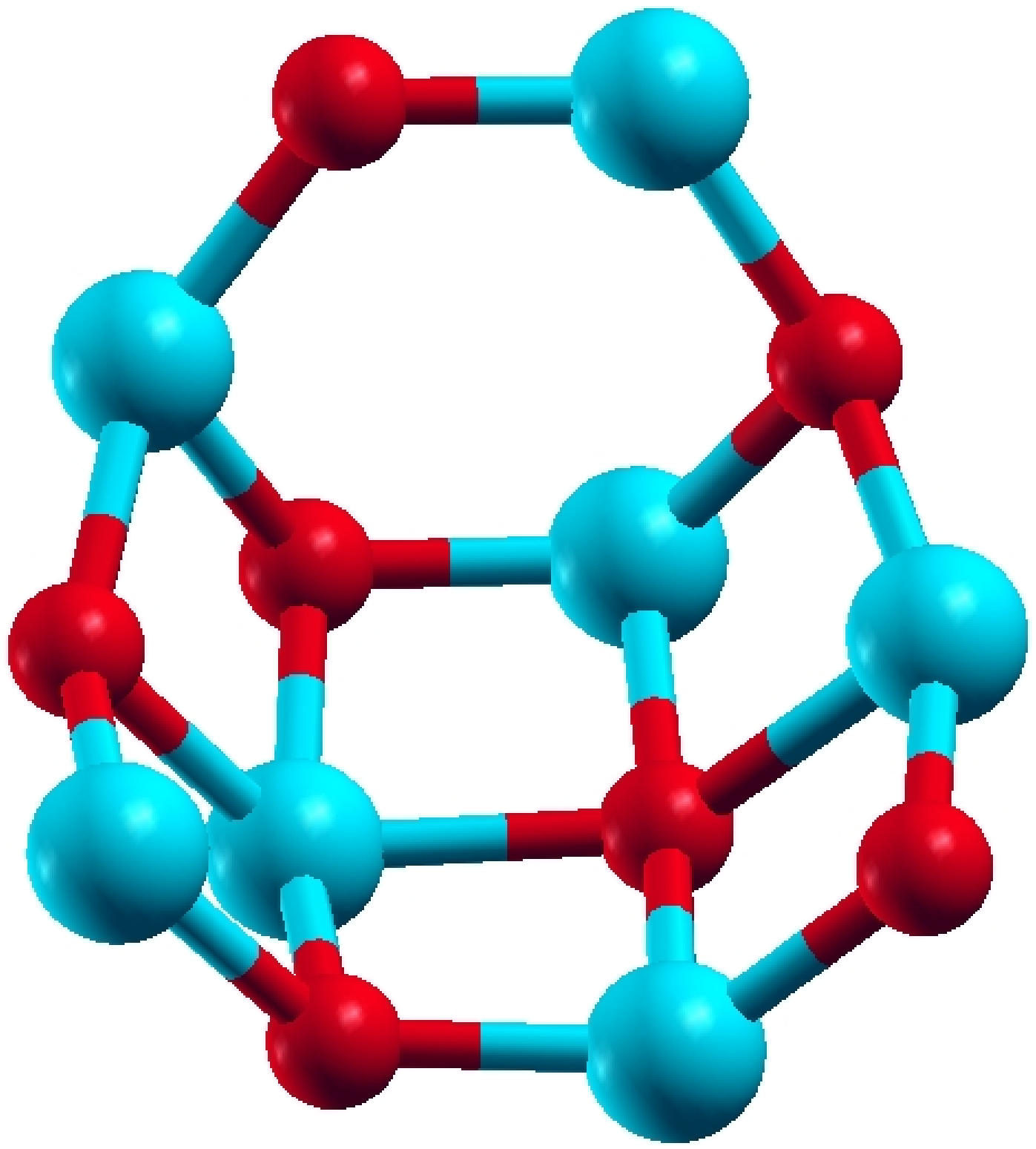} 
\includegraphics[width=5cm]{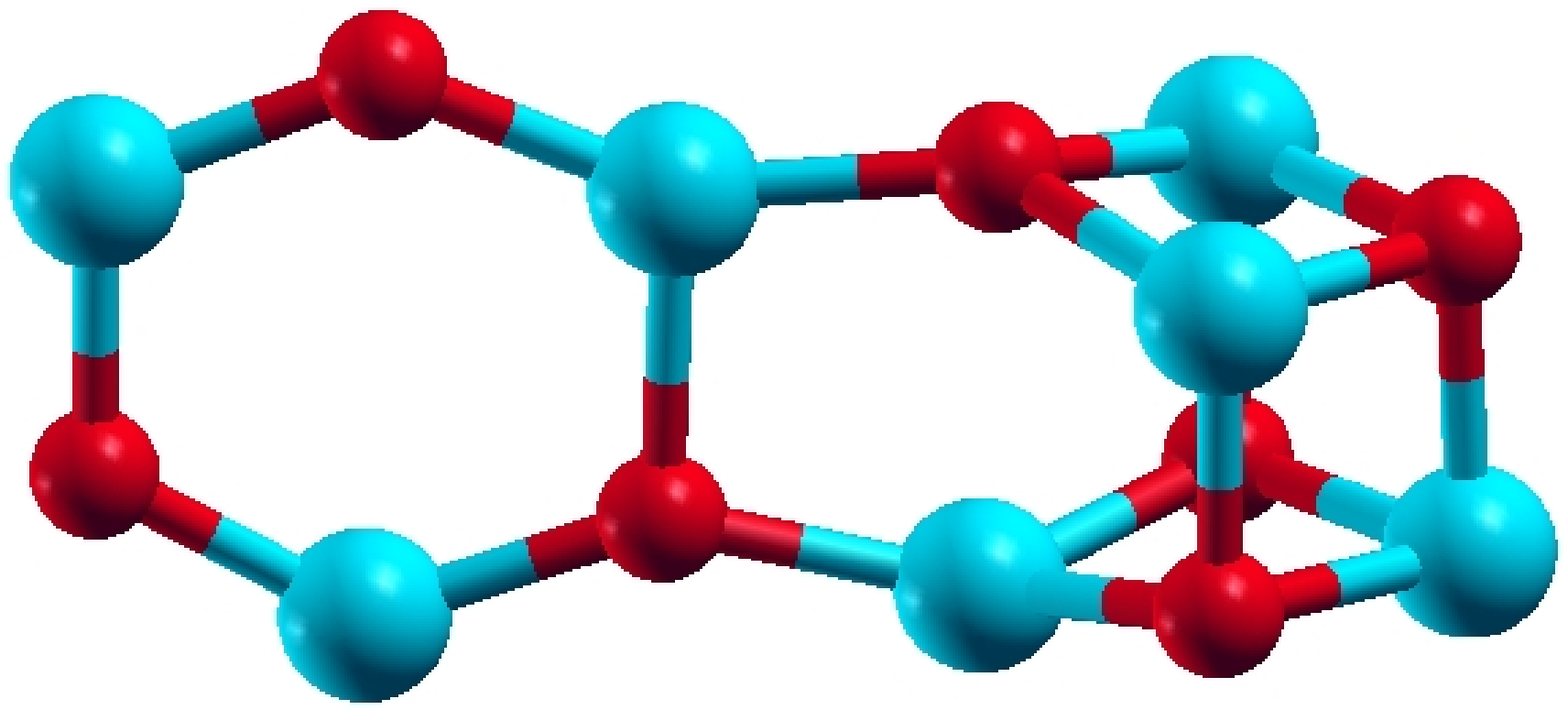}

\hspace{0cm} 7g \hspace{5cm} 7h \\
\hspace{0cm} $C_1$ \hspace{5cm} $C_s$

\label{7LiF}
\end{figure}

\clearpage
\begin{figure}
\caption{(Color online) Structures of (LiF)$_8$ clusters. For the notation,
  cf. fig. \ref{1-2-3LiF}.}

\includegraphics[width=5cm]{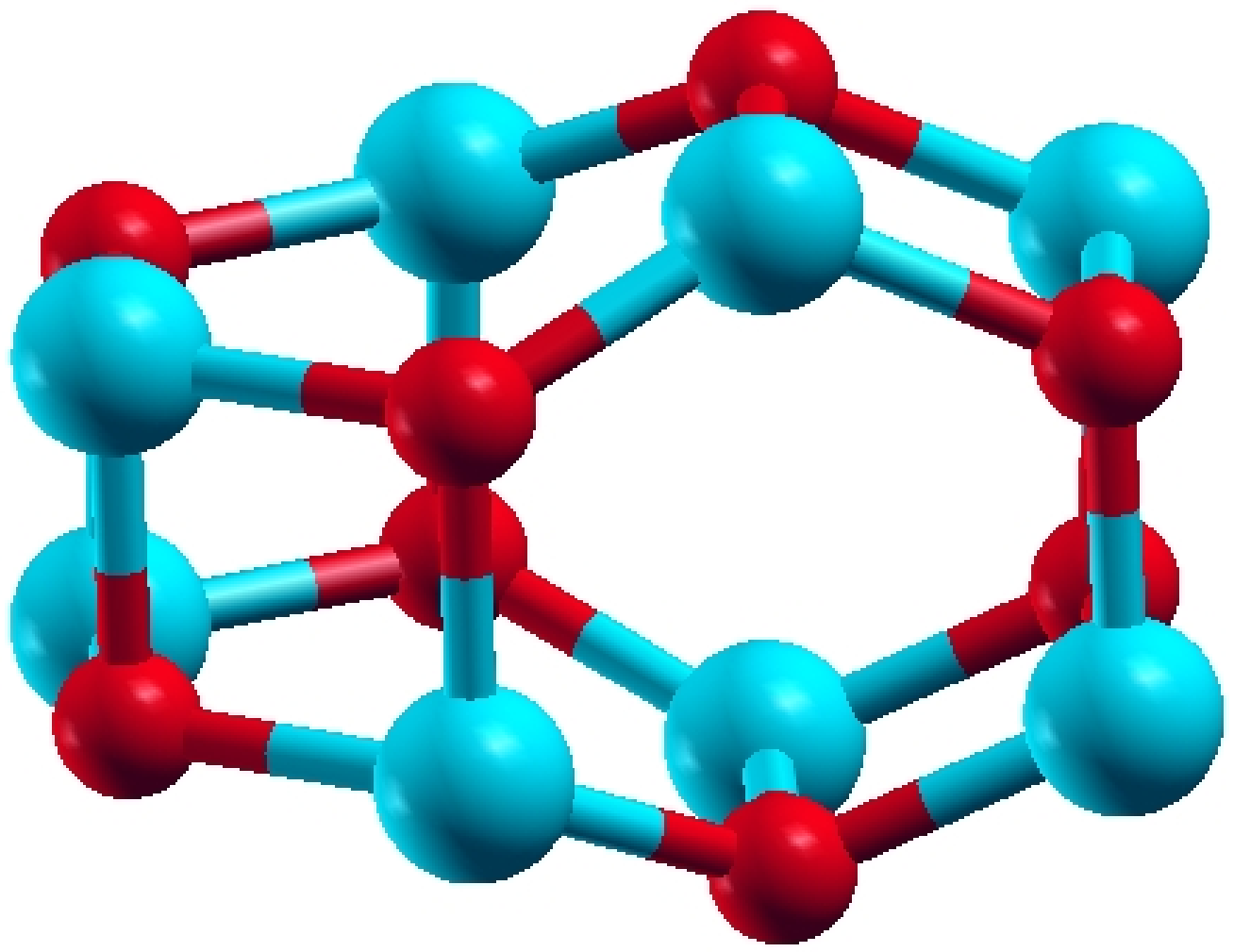} 
\includegraphics[width=5cm]{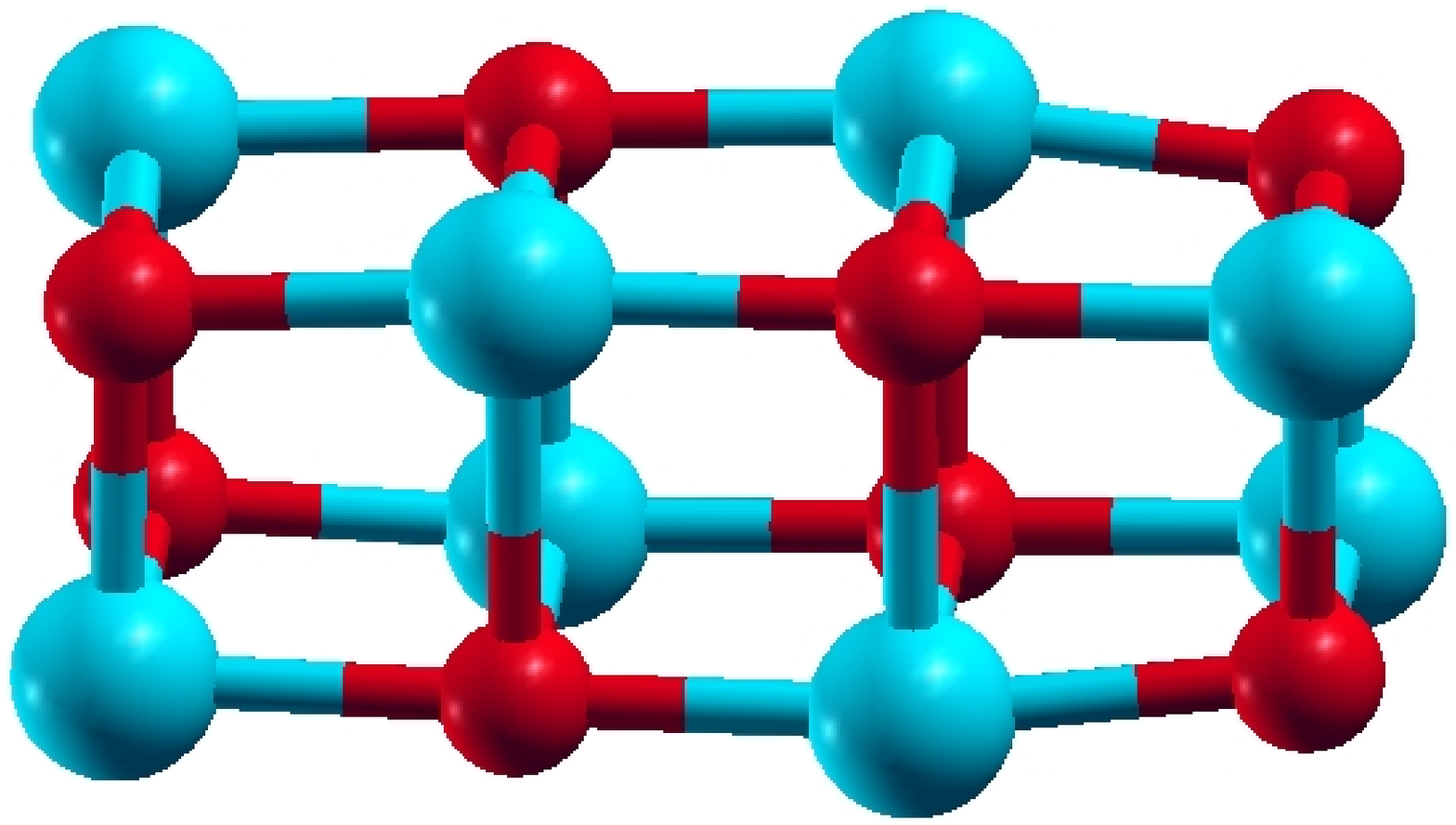} 
\includegraphics[width=5cm]{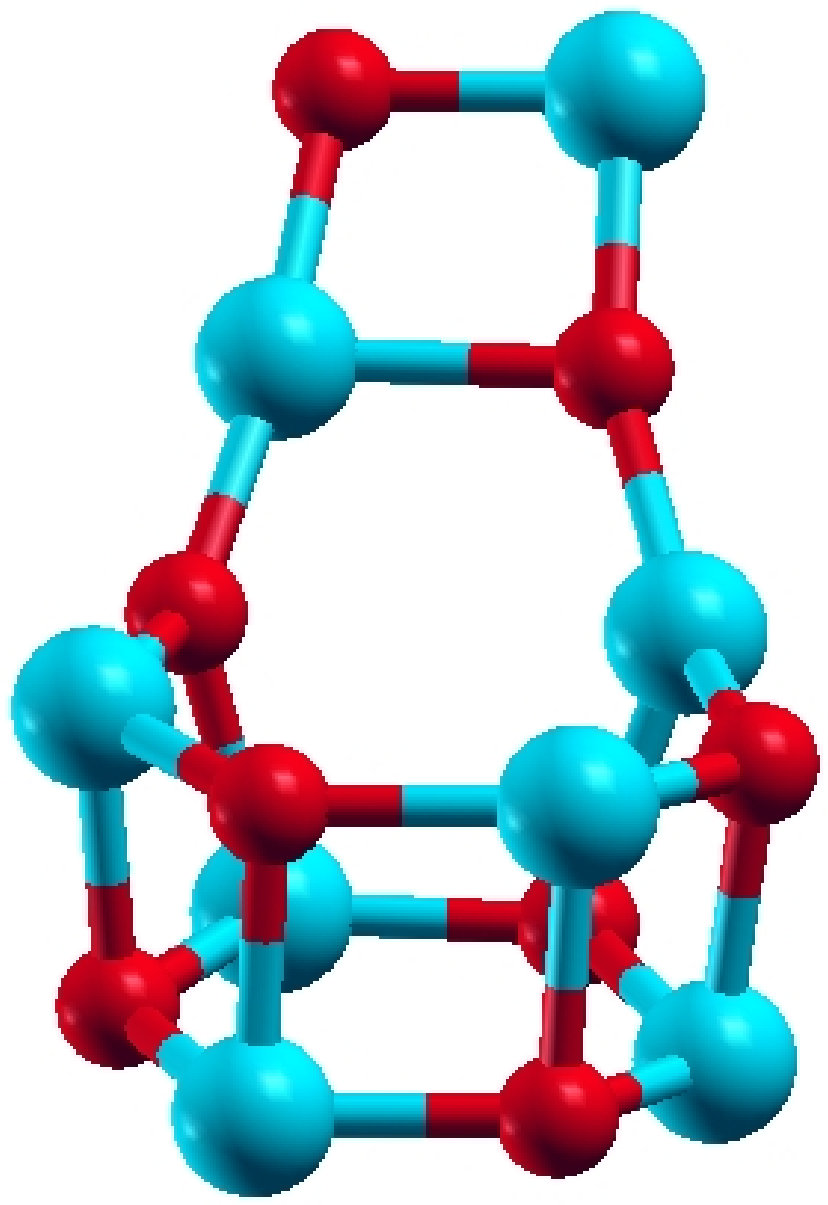} 

\hspace{0cm} 8a \hspace{5cm} 8b \hspace{4.5cm} 8c \\
\hspace{0cm} $S_4$ \hspace{5cm} $D_{2d}$ \hspace{4.5cm} $C_1$

\includegraphics[width=5cm]{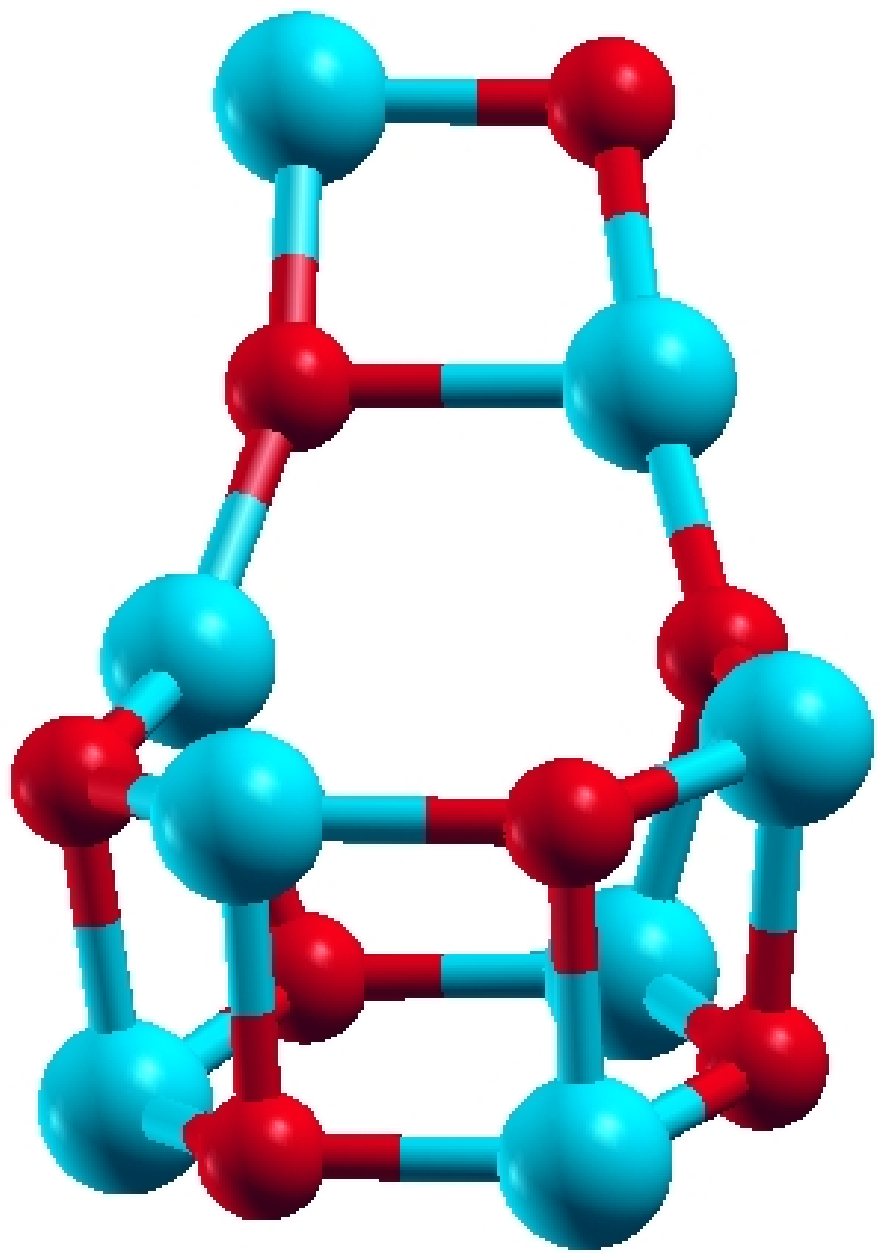}
\includegraphics[width=5cm]{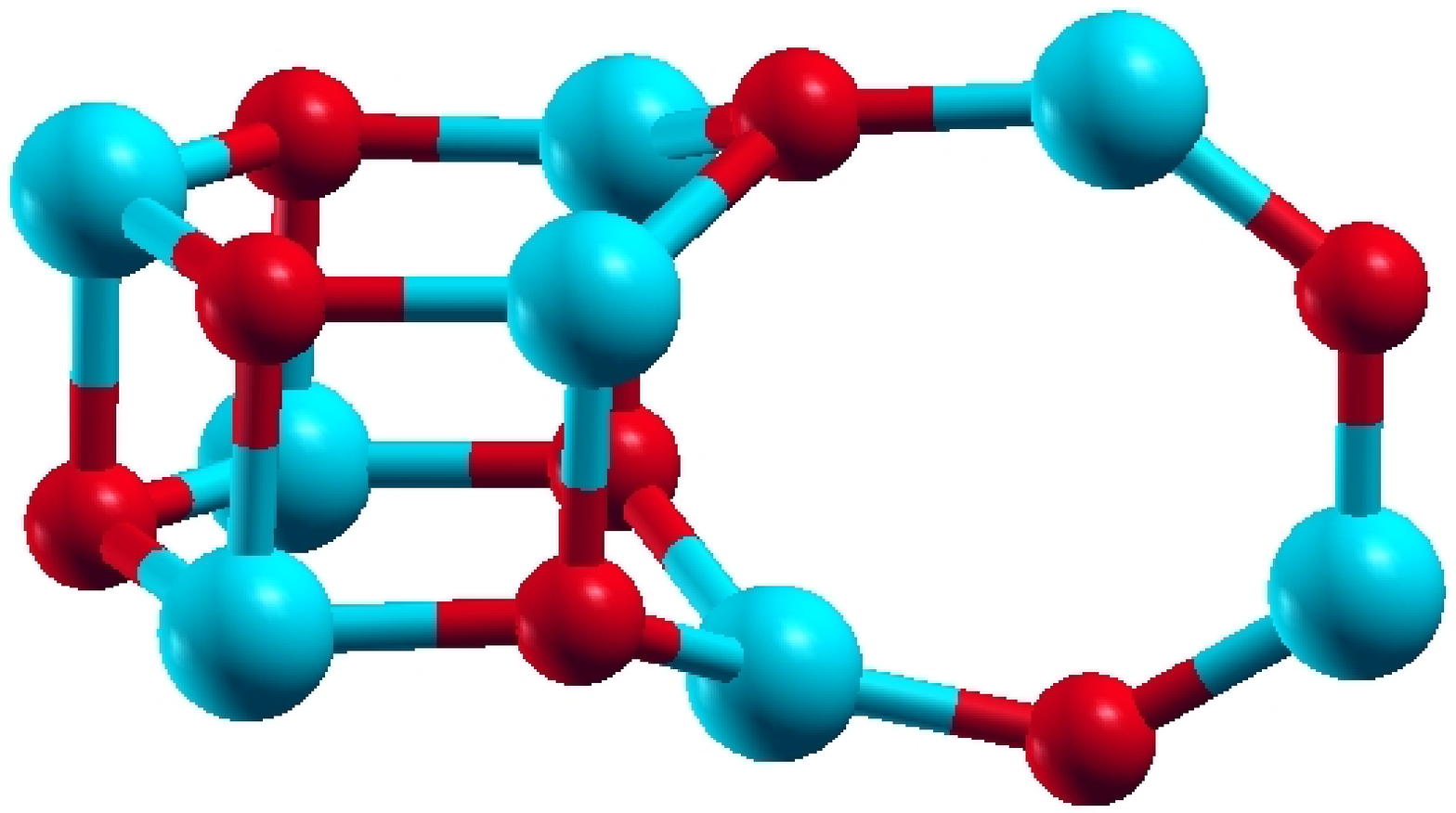} 
\includegraphics[width=5cm]{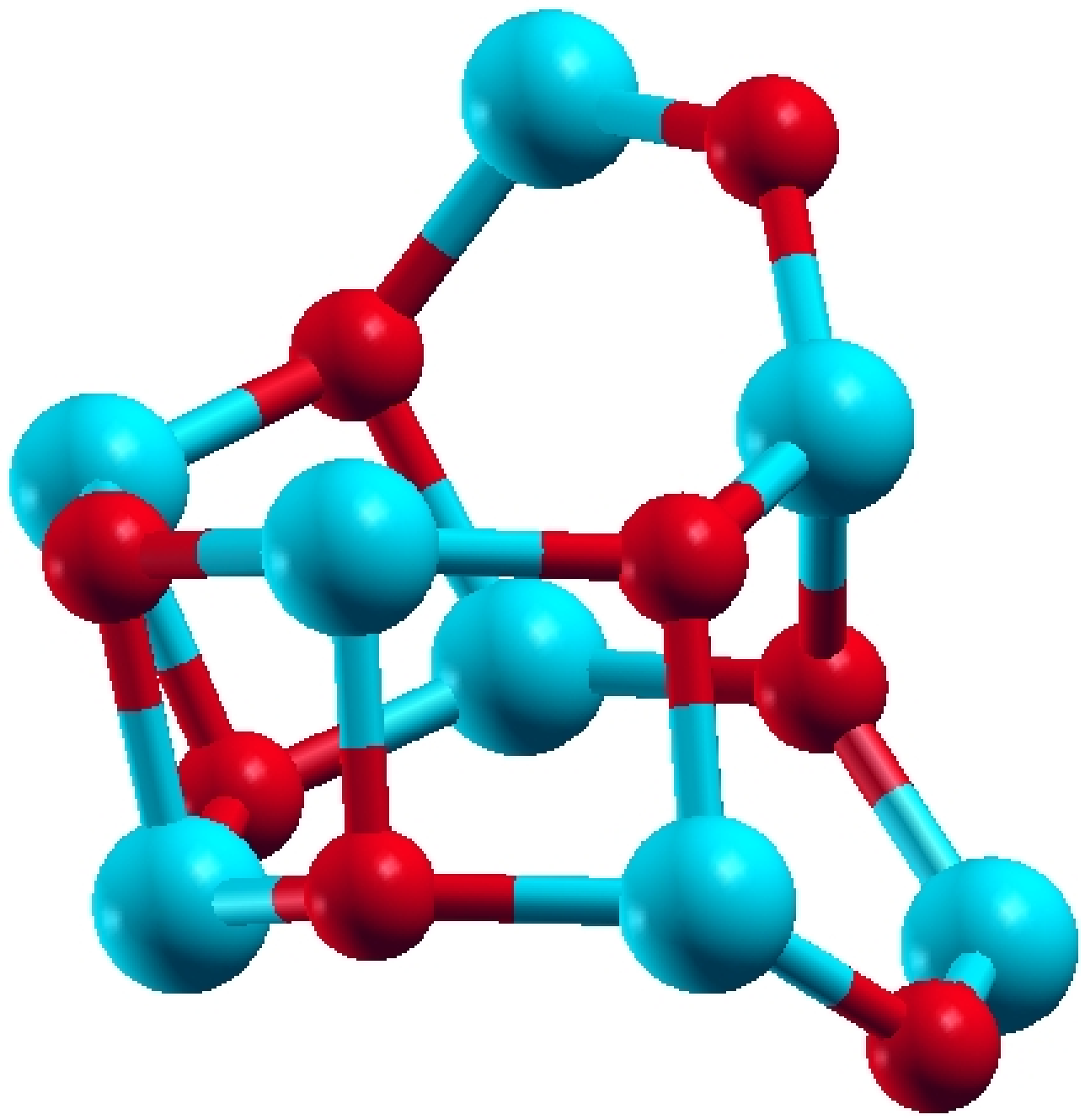} 

\hspace{0cm} 8d \hspace{5cm} 8e \hspace{4.5cm} 8f\\
\hspace{0cm} $C_1$ \hspace{5cm} $C_s$ \hspace{4.5cm} $C_1$

\includegraphics[width=5cm]{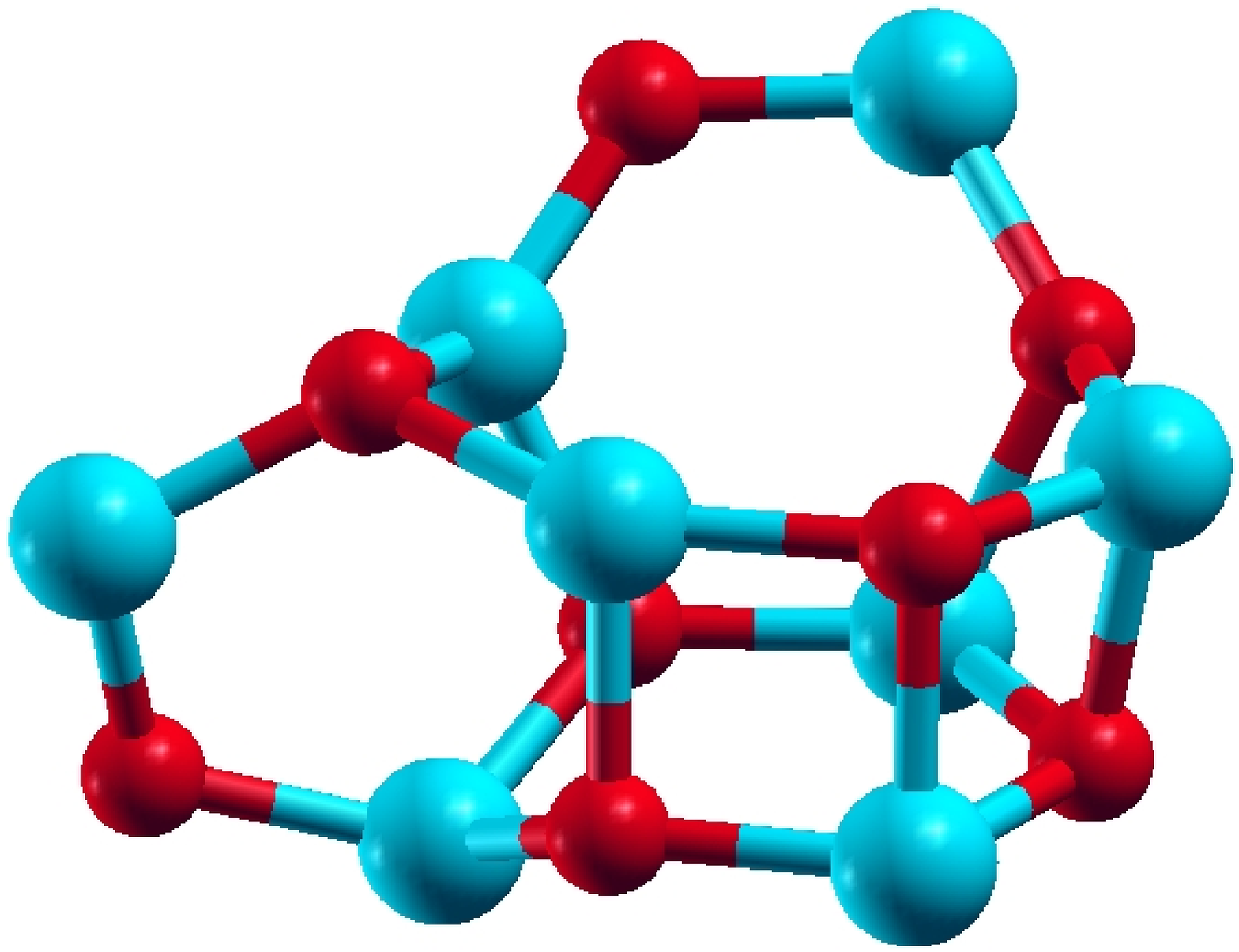}

8g\\
$C_1$

\label{8LiF}
\end{figure}

\begin{figure}
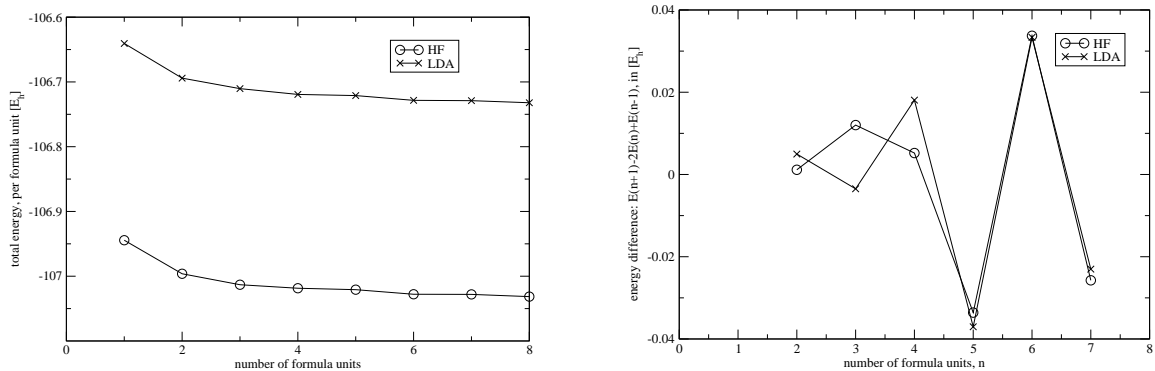

\caption{The energy (in hartree units)
of the most favorable cluster for the sizes (LiF)$_1$ to (LiF)$_8$, 
per formula unit (left), and the energy difference 
$\Delta\mu_n=E_{n+1}-2E_n+E_{n-1}$
(right). Note that if $\Delta\mu_n>0$, then
two clusters of size $n$ each are more favorable than two clusters with
size $n-1$ and $n+1$.}

\vspace{1cm}
\includegraphics[width=7cm]{energiesperfu.eps} 
\hspace{1cm}
\includegraphics[width=7cm]{differenzen.eps} 
\label{vergleichenergien}
\end{figure}

\end{widetext}

\clearpage

\begin{widetext}
\begin{figure}
\caption{Tree graph for the energy barriers of the (LiF)$_4$ cluster. 
}
\includegraphics[width=15cm]{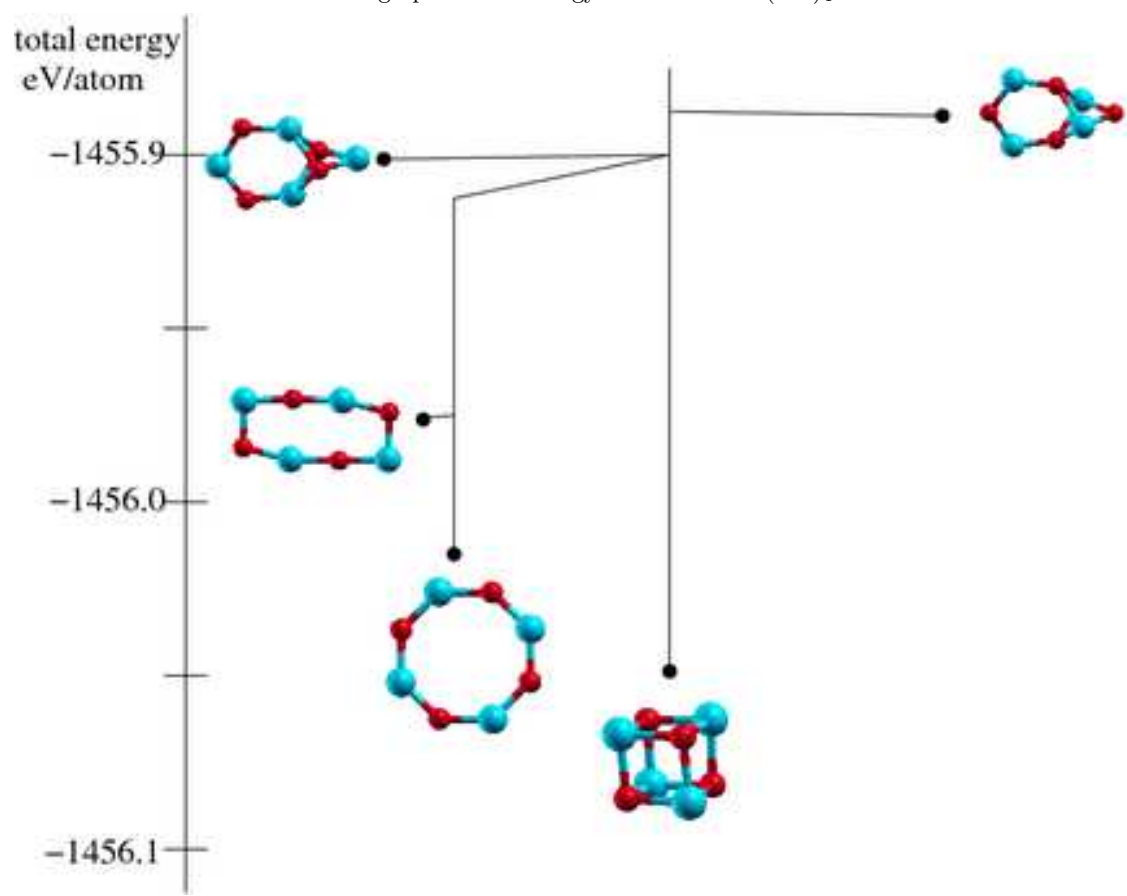} 
\label{treegraph4LiF}
\end{figure}
\end{widetext}

\clearpage

\end{document}